\documentclass[aps,prd,onecolumn,12pt,preprintnumbers]{revtex4-1}

\usepackage{hyperref}

\usepackage{afterpage}

\usepackage{float}

\usepackage{tikz}

\usepackage{graphicx}
\usepackage{epstopdf}

\begin{document}

\preprint{HDP: 18 -- 02}

\title{Banjo Drum Physics ---  sound experiments and simple acoustics demos}

\author{David Politzer}

\email[]{politzer@caltech.edu}

\homepage[]{http://www.its.caltech.edu/~politzer}

\altaffiliation{\footnotesize 452-48 Caltech, Pasadena CA 91125}

\affiliation{California Institute of Technology}

\date{June 17, 2018}

\begin{figure}[h!]
\includegraphics[width=3.5in]{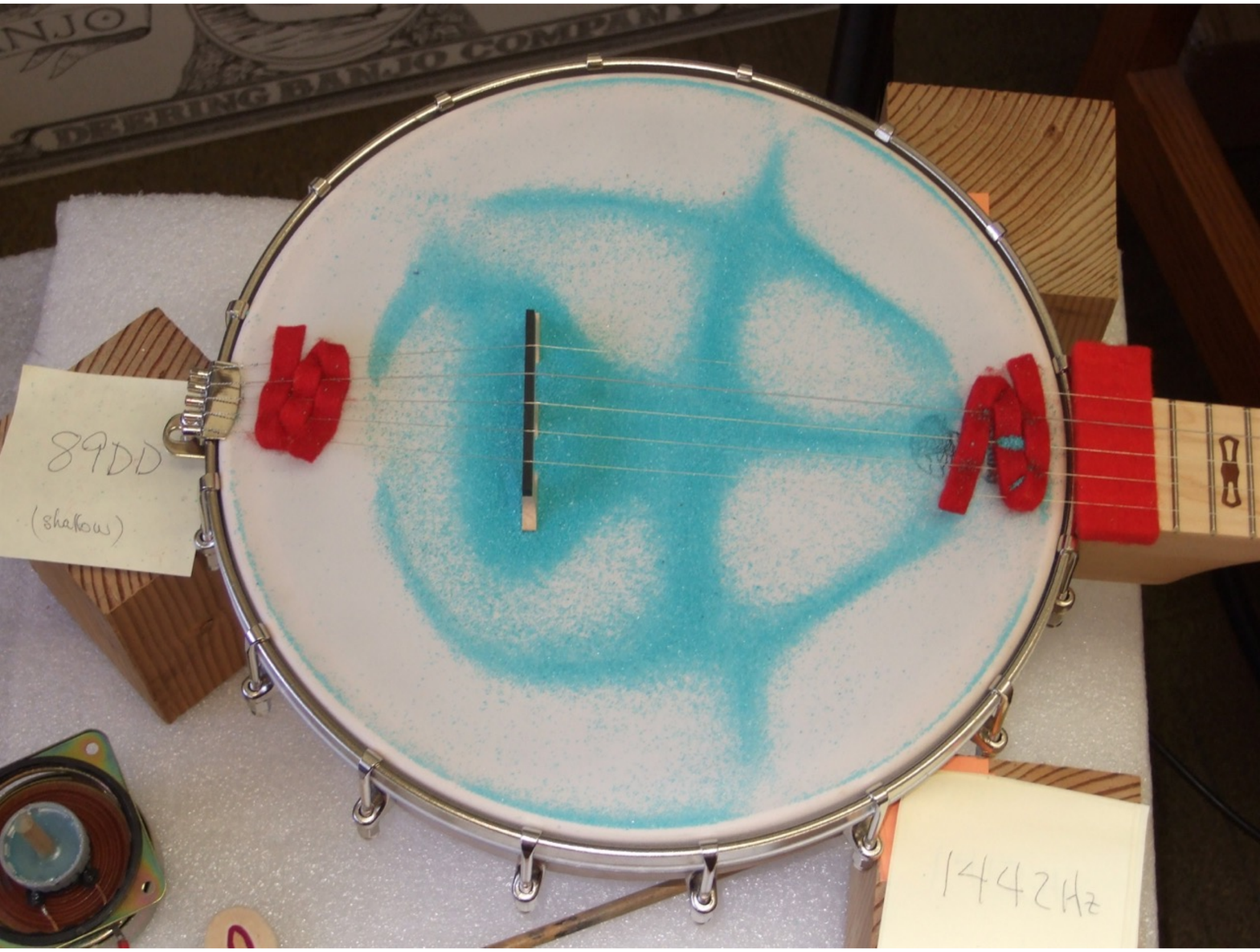}
\end{figure}

\begin{abstract}
$11''$ D mylar heads over a normal range of tensions (DrumDial 85 to 91) and ``open-back" backed pots of depths $2''$, $2{3\over4}''$, and  $5{5\over8}''$ are studied over the range 100 to 2000 Hz. Normal modes and resonant frequencies of the heads and of the pot air {\it separately} are easily identified and agree with simple expectations.  The present focus is the head $\leftrightarrow$ pot air interaction.  There is no ``gold-plated" example of a pair of head--air interacting modes that are distant in frequency from all others.  (Had there been such a pair, their interaction could have been isolated and studied in detail.)  Nevertheless, there are a few cases where there are hints of the kind of interactions expected from a simple theory.  The investigations also offer several examples of banjo physics, including aspects of bridge position and rim flexibility, and some dramatic examples of the perils of sound recording, including floor bounce and room sound.
\end{abstract}

\maketitle{\centerline{\large \bf Banjo Drum Physics ---}

\centerline{\large \bf sound experiments and simple acoustics demos}

\section{Introduction}

Strings $\rightarrow$ bridge $\rightarrow$ head $\rightarrow$ sound.   That's the basic storyline of banjo sound and the essence of $\cal BANJO$.  It's what all banjos have in common and what makes them immediately identifiable.  Differences and distinctions between instruments are subtle and often elude the uninitiate.  However, virtually every design variation ever implemented in  nearly four hundred years in North America has its enthusiasts and is still in production (although some more limited than others).  Interest in the possible variations is fueled by the many adjustable and swappable parts and by their being not too hard to make.  Also, many players imagine that hardware is a key  difference between their own performance and that of someone they admire.

Often, a productive approach to a better understanding of the physics is to investigate sub-systems and how they work.  One then considers how those systems work together to make the whole.
Strings are an example of sub-system whose isolated behavior is simple to understand --- at least roughly --- and whose coupling to the rest of the instrument is weak.  ``Weak," here, means that the strings can vibrate through a great many cycles after being plucked before their motion changes appreciably from its isolated form.  In general, the motions of sub-systems that are weakly interacting with the whole are recognizable cousins of isolated versions of those sub-systems.  Sometimes, clearly distinguishable sub-systems interact so strongly that their individual, simplest motions are lost.  In the case of a tone ring sitting on a wood rim, the interaction is typically so strong that the two simply move together, at least for their lowest frequency motions.\cite{ring-ring}  And sometimes, the interaction, no matter how weak, is essentially non-linear.  In such cases, the interaction between the sub-systems introduces some behavior that is new in a very essential way.  (That contrasts with the prosaic behavior of linear systems which pass vibrations from one to the next {\it via} some linear filtering.)  Many such non-linearities are instances of ``parametric" coupling.\cite{Rayleigh}  String-stretching by the floating bridge is an example.\cite{FM}

The pot air -- head sub-system seemed promising because each part can be isolated in an experimental set-up;  by themselves, they're  well-understood; and their interaction has a relatively small impact on the head motion.  The geometry of the pot design defines the possible air motions inside and is known from experience to be an important distinguishing feature of different banjos.  The motions of the head and the air {\it separately} can be approximated by  classic, idealized, solved physics problems.  Their interaction already drew the attention of Rayleigh (as a drum, not a banjo[!]).\cite{Rayleigh}  He noticed that the kettle of the kettledrum helped produce a definite pitch, at least for particular strikes.  Others then followed, well into the $20^{\text{th}}$ Century.  The correct equations, at least as linear, first approximations, were never in doubt.  The issue has always been extracting properties of their solutions.  Following Rayleigh, kettledrum researchers have mostly focused on precise determination of frequencies of low-lying harmonics and/or gross features of the radiated air motion.  Sophisticated  computer calculations became an essential part of this endeavor.\cite{timpani}  In contrast, my goal is to get a better, overall, qualitative understanding of the banjo.  To be honest, the potential for impact on design decisions is remote.  At best, it might inspire some choices.  But physics has never really played an important role in the development of musical instruments.  That's in no small part because people don't even agree on what they hear in a given instance.  Furthermore, no one knows how to turn what people hear into quantified descriptors --- at least not at the level of discrimination important to music lovers.  Inspiration, trial-and-error, and Darwinian evolution are the prime movers of design development.

In an earlier paper\cite{theory}, I laid out the simplest picture I could imagine to account for the pot air -- head interaction.  I then set out to isolate some example where one could see the basic processes in action.  I hoped to find something so simple that a few measurements and a bit of algebra would be convincing evidence of the validity of the more general picture.  That was done for the Helmholtz resonance a long time ago with the guitar --- and with the banjo.\cite{cylinders}  In that case, the air resonance is an overall expansion and contraction that pushes some air in and out of the pot (through the flange or out the back).  That couples to the overall up-and-down mode of the soundboard or head.  These are typically the lowest modes of the body of a string instrument.  (The Helmholtz resonance also couples more weakly to any higher soundboard mode that involves a net flow of air in and then out.) 

In the meantime, I came across a very respectable source substantiating an approximation that I described and viewed as essential in my theoretical proposal.\cite{theory}  In the middle of reviewing modern work on timpani and detailing the author's own approach, ref.~\cite{kergomard-book} notes that, for the range of parameters present in their drums, the full-numerical computer evaluations agree fairly well with a much simpler calculation which is equivalent to what I described in ref.~\cite{theory} as first-order back-reaction.  The approximation rests on the observation that the air has only a small effect on the drum head motion.  Its validity depends on the actual values of the physical parameters.  And it is certainly only more so for the banjo than the kettledrum --- because the head is relatively tighter and the enclosed air volume is much shallower.

For my own experimental endeavor, I ultimately stripped three banjos down to rim and head --- no bridge, no strings.  That's the simplest possible version of head and enclosed air.  The circular and cylindrical geometries make them good candidates for modeling by the ideal systems solved in physics textbooks.  The experimental variables were head tension (within normal playing range) and the three rim heights.\cite{sound-files}  Excitation could be piano hammer head taps or electronically driven piezo disk at various locations.  Microphone location was varied.  I limited my focus to 100 to 2000 Hz because that range contains a limited number of head and air modes that could be unambiguously identified using only simple tools.  (As frequency increases, two things happen: the resonances get closer and closer together in frequency, and their order [at least for the drum head] need not be the same as it is for the idealized system, which makes it harder to know what spatial shape goes with any particular frequency.)

\subsection{Theory Summary Review}

The general, linear analysis of the head -- air system (as performed in the frequency domain, as opposed to directly in time) involves all of the head normal modes and all of the air normal modes.  The added perspective of the ``theory" I offered is a way to think about which modes are most important to a given part of the problem.  As noted in ref.~\cite{theory}, one can use one of the standard representations of the Green's functions as a guide.  In particular, there is a sum over projections onto and out of an outer product of normal modes.  (Whew!)  The hope was to find an air and head pair which interacted more with each other than with any other modes.  In such a case, the variation of the observables with the controllable parameters could serve as simple support for the basic notions.

\subsection{Conclusions --- the short version}

There is no such pair --- at least for the range of design parameters typical for the banjo.  A pair would have been singled out as having a particularly strong connection to each other (i.e., relative to the influence of other modes) if their frequencies were near each other and their spatial structures were effective at producing such interaction.  The latter requires that they push at roughly the same places at the same time.  The opposite would be a case if, over a region where one pushes at a given time, the other pushes in one part of the region and pulls equally effectively in another.  The net effects would cancel.  The pair of modes would neither be moving together or in opposition.

What one does see is that there is definitely an overall effect of the pot air on the drum motion that depends on the geometry of the pot and the tension of the head.  Some of that dependence makes qualitative sense, but the details reflect the net effect of many influences at once.

Had there been a single pair of closely related modes, their frequency would likely have stood out in normal playing ---  either noticeably strong or noticeably suppressed.  But that's precisely what we don't want in a musical instrument.  I once read that what is desired is ``reverberant rather than resonant."  We want an instrument's response and timbre to change very slowly (if at all) in going from note to note.  An isolated resonance produces undesirable sound by singling out one note or frequency interval.  The presence of many complex and competing factors smooths things out.

I did identify a few cases where the resonances of the combined systems are at least suggestive of the interaction between head and pot air.  However, none of these seem clean enough to warrant further detailed study.

\section{outline}

Absent any simple or compelling examples, the techniques, asides, and tangents may well be of greater interest and value than the central ``results."  So no obvious, logical order of presentation suggests itself.  Here is an outline of what there is.

String tension ({\it via} the bridge) distorts the shapes of the head resonances, as shown in the Chadni figures of \S III.  This dramatically alters which mode can talk effectively to which other mode.  The focus on the nearly perfectly circular drum is really only a step on a route to understanding the drum aspect of the banjo.

\S IV presents the measured frequency spectra for the head alone and as mounted on the three rims (with a back that simulates openback banjo playing).\cite{openback}  The spatial structures of the head modes are identified.  These identifications show that nearly every air mode has a spatial structure that precludes its talking to the head modes  nearest in frequency.  Only a couple of cases are even vaguely in accord with how the air might be expected to influence the head.

\S V discusses how head modes were identified.  Chadni figures reveal the head mode spatial structure that goes with each resonant frequency.  For the air modes of a cylindrical cavity, simple theory and measurements were previously shown to agree quite well.\cite{cylinders}

\S VI describes an example of an impact (albeit tiny) of rim stiffness on produced sound.  The particular observed effect is likely too small to be noticeable in actual playing.

Some of the particular aspects of actually doing the measurements are collected in \S VII.  That includes (A) impact of microphone positioning on the measured spectra, e.g., room sound and floor bounce; (B) other lessons from  head taps, specifically the relevance of bridge position and reproducibility of measurements; (C) some details of DrumDials; and (D) a consistency check of the apparatus and software.

A concluding \S VIII offers reminders that spectra do not tell the whole story of what we hear, even were they to be measured with absolute precision.

\section{Head modes with damped strings}

\begin{figure}[h!]
\includegraphics[width=3.0in]{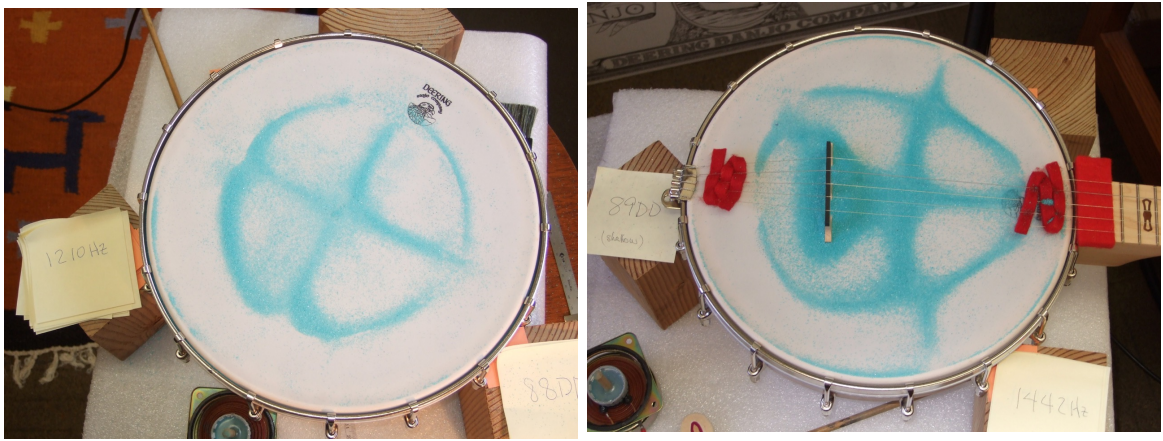}  \includegraphics[width=3.0in]{89DD-2-2-Chladni.pdf}
\caption{Chladni plate node lines, without and with bridge and tuned, damped strings}
\end{figure}

The circular symmetry of the pot without strings limits which head and pot air modes can talk to each other and exchange energy.  In theory (if the simple idealization is taken literally), head and air modes must have the same number of ``azimuthal" (i.e., diameter) node lines or planes if they are to influence each other's motion.  FIG.~1 shows analogous modes without and with the bridge and strings.  ``Analogous" means that one would smoothly transition to the other if the down pressure of the bridge were smoothly varied from zero to its normal value.  And there is a one-to-one map of circular drum modes onto the distorted ones.  Distinguishing features of the modes in the circular case are the number of diameter lines and the number of concentric circles.  (They are traditionally labeled by those two integers).  Turning on the bridge down-pressure distorts those lines continuously.  Consequently, the head in the assembled banjo can excite many more air modes than it can in the symmetric, round drum configuration.  And each excited air mode can react back on many more head modes.  For a real banjo, matching the number of azimuthal node lines might not be a significant criterion for back-reaction.  However, proximity in frequency is always relevant.

A published, professional effort to map head mode shapes using laser interferometry suggests somewhat less severe distortion due to the bridge and strings.\cite{stephey-moore}  Getting the sand to dance for the Chladni figures shown here required rather large amplitude vibrations.  (It was loud enough to elicit the first banjo noise complaint ever from a neighboring office.)  In the laser interferometry experiment, the head was driven acoustically by a speaker located 1 meter away.  Presumably, that speaker's sound volume was not beyond normal music loudness.  Also, such a  speaker's pressure wave as it arrives at the head is a very poor spatial distribution match to any but the lowest head mode.  So the banjo head induced amplitude would be a $3^{\text{rd}}$ order effect, i.e., much, much softer than normal playing.  The Chaldni figures shown here were induced mechanically.  (See \S V.A.)  So the differences in node patterns could well be attributed to non-linear effects in the sand case.  At higher amplitudes, non-linear effects due to head stretching and stiffness might be relevant.  (At these loudness levels, the air motion remains quite linear.)  The sand patterns were properties of a steady-state, constant amplitude rather than a transient from a plucked string.  Certainly a plucked string sound dies off {\it eventually} to something very soft and linear.  But an individual pluck, going from its initial attack to when it dies out, can be heard and measured over a range of about 50 dB.\cite{stephey-moore}  So {\large \bf loud} may be more appropriate for the generation of the characteristic pluck sound timbre.  --- Alternatively, the linear head modes might, indeed, be significantly distorted by the bridge and strings, even at small amplitudes.

Several details of FIG.~1 deserve comment.  Sand accumulates where the head hardly moves.  Those are the node lines.  However, the bridge depresses the head.  So sand migrates to the base of the bridge whether it's moving or not.  The two images in FIG.~1 were taken on different occasions.  The DrumDial head tension settings were 88 and 89, respectively (as noted in the photos).  That corresponds to a nominal increase in tension of about 10\%.  (See \S VII.B for a DrumDial discussion.)  For an ideal, circular membrane, the frequency is proportional to the square root of the tension.   The observed frequencies differ by somewhat more than that but might be due to the uncertainties in the tension (from the steepness in tension of the DrumDial and uncertainties in its reading), the extra restoring force on drum motion from the damped strings, and/or the settling of the head during the course of a long series of measurements.

\section{100 to 2000 Hertz, with and without the pot}

Frequency spectra are compared in FIG.~2.   The sound was produced by multiple taps on the head.  Subsequent sections below describe how it was done and how the mode shapes corresponding to the particular peaks were identified.  

The first, ``head only," spectrum is for a standard mylar head tightened to a reading of 85 on a DrumDial (details in \S VII.B).  85 is about the lowest tension reasonable for a mylar head for normal playing.  The low head tension is of particular interest here because a higher tension head would move less air and be less influenced by the motion of that air.  And it is that back reaction of the air back onto the head that I wanted to investigate.  91 is about the highest tension available with a standard mylar head.  It is particularly useful when identifying the relation between observed, ``head only" peaks and the theory of an ideal drum head.  The high tension makes its motion less sensitive to other forces, e.g., its own stiffness, air loading, and pressure variations inside the pot.

The DrumDial calibration presented in \S VII.B suggests that all peak head frequencies at 85DD should be about 0.74 times their 91DD counterparts.   (Those frequencies scale like the square root of the tension, and that works out pretty well.)

The three different rims (heights $2''$, $2{3 \over 4}''$, and $5{5 \over 8}''$) are backed with an open-back playing simulator (aka the ``belly back").\cite{openback}  They have identical, standard heads, all tightened to the same tension as the accompanying ``head only" curve.

\begin{figure}[h!]
\includegraphics[width=6.0in]{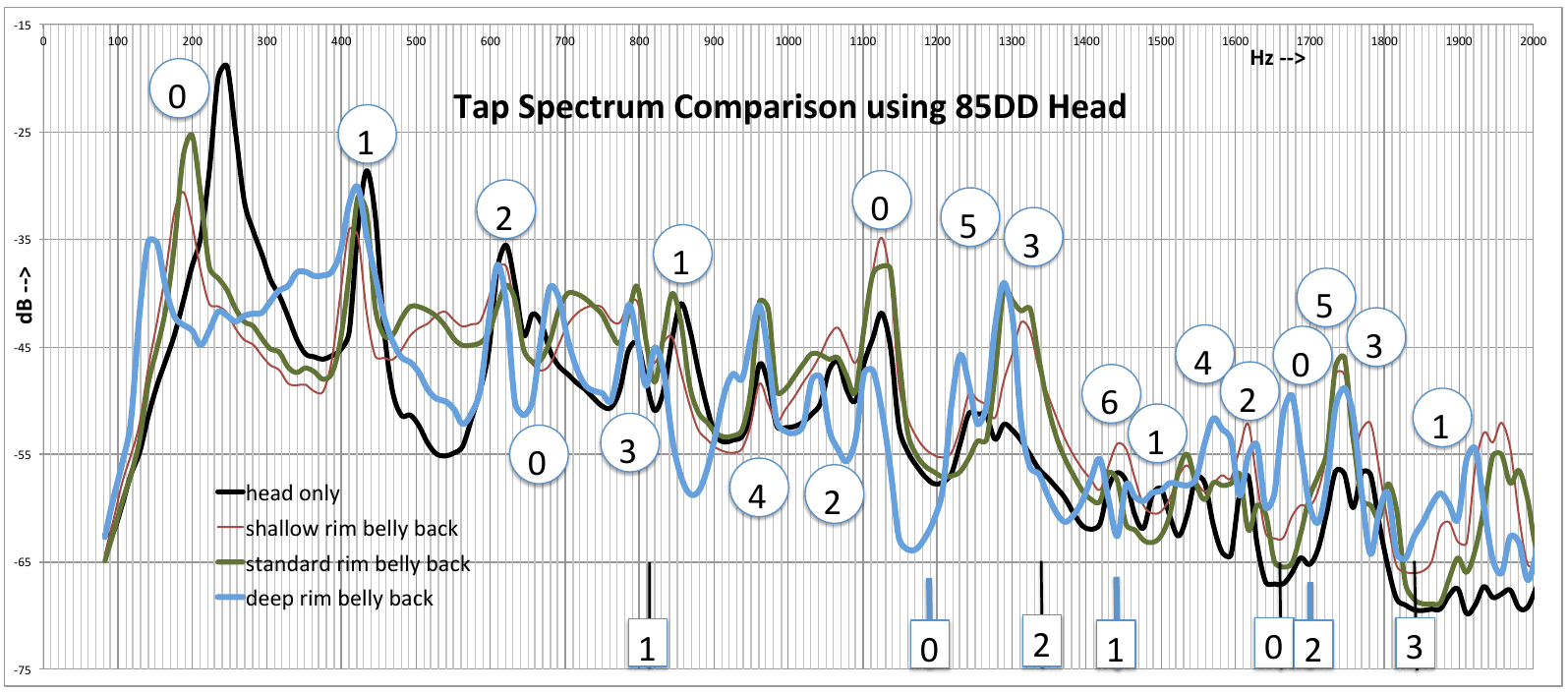}
\includegraphics[width=6.0in]{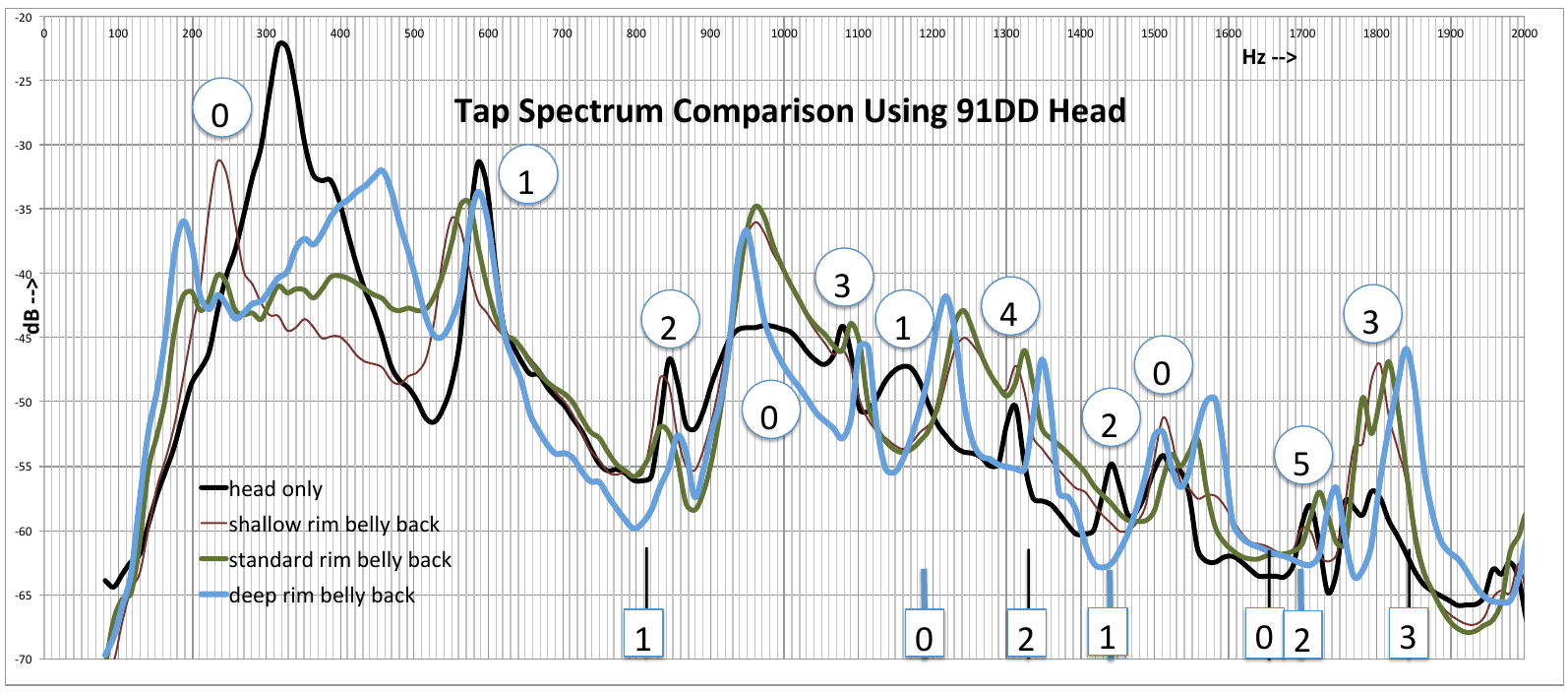}
\caption{Spectra of heads at 85 DD and 91DD,  solo and mounted on three rims (open-back backed); the black vertical lines at the bottom mark the air modes common to all rims; the shorter, thicker lines mark the additional air modes present in the deep pot.  Circles label the number of diameter node lines for the solo head; squares label air mode diameter node lines.}
\end{figure}

The vertical lines at the bottom of the graph are the air mode frequencies calculated for cylindrical cavities of those dimensions.   Previous measurements performed on these same rims (with the head and back replaced by ${3 \over 4}''$ plywood) produced frequency peaks at just those frequencies.\cite{cylinders}  The four longer lines are common to all three rims.  Those frequencies only depend on the common rim diameter because the air pressure is constant in the squat direction, i.e., perpendicular to the head.  The three shorter lines are additional resonant frequencies that arise for the deep pot due to an additional variation in that squat direction.  (Analogous resonances for the shallower pots appear at yet higher frequencies.)

The lowest peaks in both graphs reflect the well-understood interaction of the lowest head mode with the Helmholtz resonance of the cavity.\cite{cylinders}  (The frequency of the latter decreases with increasing pot volume, but it is also very sensitive to the size of the opening to the outside air.  Having not been as careful with that adjustment here as in earlier experiments, I am not concerned by the $2''$ and $2{3 \over 4}''$ rims appearing slightly out of order in the 85DD plots.)  This lowest head mode motion is simply overall up and down, with no nodes except for the stationary edge at the rim.  In standard notation, it is the (0,1) mode --- for zero diameter node lines and one circular node line (i.e., at the fixed edge).  The Helmholz mode is an overall expansion and contraction of the air in the pot, which pushes air in and out its vent.

In the exactly circular, idealized, soluble model, head and air modes with different diameter node numbers do not interact.  When followed over a full cycle of the driving frequency, their net effect on each other is zero.  In contrast, the number of nodal circles is not crucial because there never is a close spatial match of the head to air modes.  That is because the outer edge of the head is always fixed (i.e., is a node line for every mode), while the outer edge of the cylindrical pot is always an ``anti-node," i.e., an oscillating maximum of pressure.  So in FIG.~2, the number of diameter node lines is included as a label on the various peaks.  (With the crude experimental techniques employed here, the identifications for the 85DD head modes above 1650 Hz are a bit dicey.)

If there is a head resonance with no nearby air resonance with matching diameter node number, all three rims are expected to have peaks at that head frequency.  Depending on how strictly you interpret that as a prediction, you can find eight or so examples in FIG.~2.  (There are a few counterexamples, too.)

We should expect the deep pot air modes to have greater influence on the head than the shallower two because there is more air moving at that frequency.  The deep mode has more inertia and, therefore, a stronger return force for a given frequency.  The deep pot air mode around 1700 Hz might be very relevant to the deep combined modes around 1630 and 1660 Hz, distinguishing the deep rim in that frequency range from the others.

Perhaps the most suggestive case of head -- air interaction is around 1800 Hz for the 91DD head.  All rims have an air resonance around 1840 Hz.  It ``pulls" the combined resonance frequencies up from the 1800 Hz head resonance, but it does so most effectively for the deep pot -- again, a triumph of inertia.  The 1655 Hz air common air resonance might be doing the same sort of thing to the head resonance at 1530 Hz.

With the 85DD head, the three rim resonances line up around 785 Hz, presumably driven by the corresponding 3 node head resonance.  The very nearby 1-node air resonance doesn't talk to those, but it does seem to effect the combined modes that originate with the 860 Hz head mode.  Again, the deep pot air has the strongest interaction.

All in all, that's pretty disappointing from the perspective of the original goal.  However, it's pretty enlightening in terms of how complex the interactions are, at least from the perspective of normal modes.

\section{Identification of head modes \& making Chladni figures}

Identifying the spatial structure of resonant modes is a crucial step in understanding how they interact.  The measured cylinder air modes agree so well with calculated frequencies that there is no reason to doubt the implied spatial structures.  Heads are potentially a different story.  The simple theory is for membrane vibrations where the only relevant force is a constant, uniform tension.  That ignores the inherent stiffness of the head material (which resists stretch, bend, and shear), the inertia of the air that has to be moved as the head moves, and the springiness of that air as it gets compressed and ``rarified."  So it is reasonable to try to observe the spatial structure of the head modes.

The first obstacle was that banjo heads do not typically exist in isolation, at least not under tension.  The tension is usually provided by stretching the head over a rim.  The shallowest rim I had was $2''$ deep.  I was able to convince myself that $2''$ of rim, if left totally open otherwise (i.e., with no bottom to further enclose of the cylindrical volume), is pretty close to no rim at all.  The key item was a $10''$ 90DD {\it pre-tensioned} mylar drum head, pictured in FIG.~3.  Its spectra with and without a $2''$ cardboard rim are plotted in FIG.~4.

\begin{figure}[h!]
\includegraphics[width=3.7in]{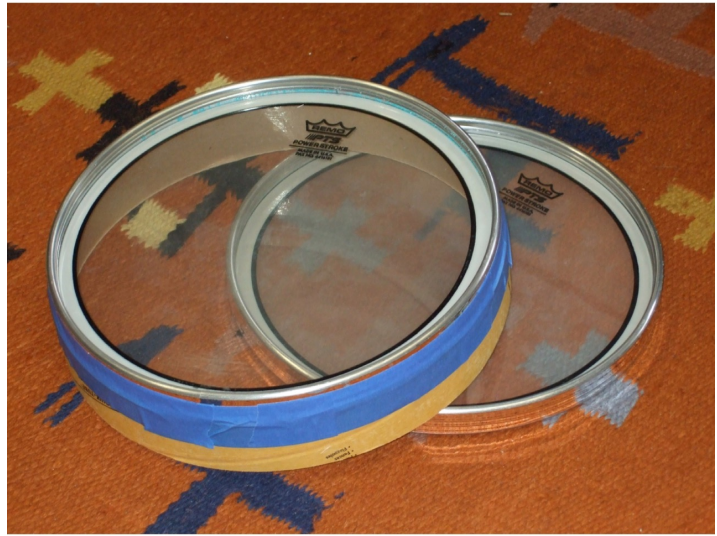}
\caption{$10''$ pretensioned drum heads, with and without a $2''$ rim}
\end{figure}

\begin{figure}[h!]
\includegraphics[width=6.0in]{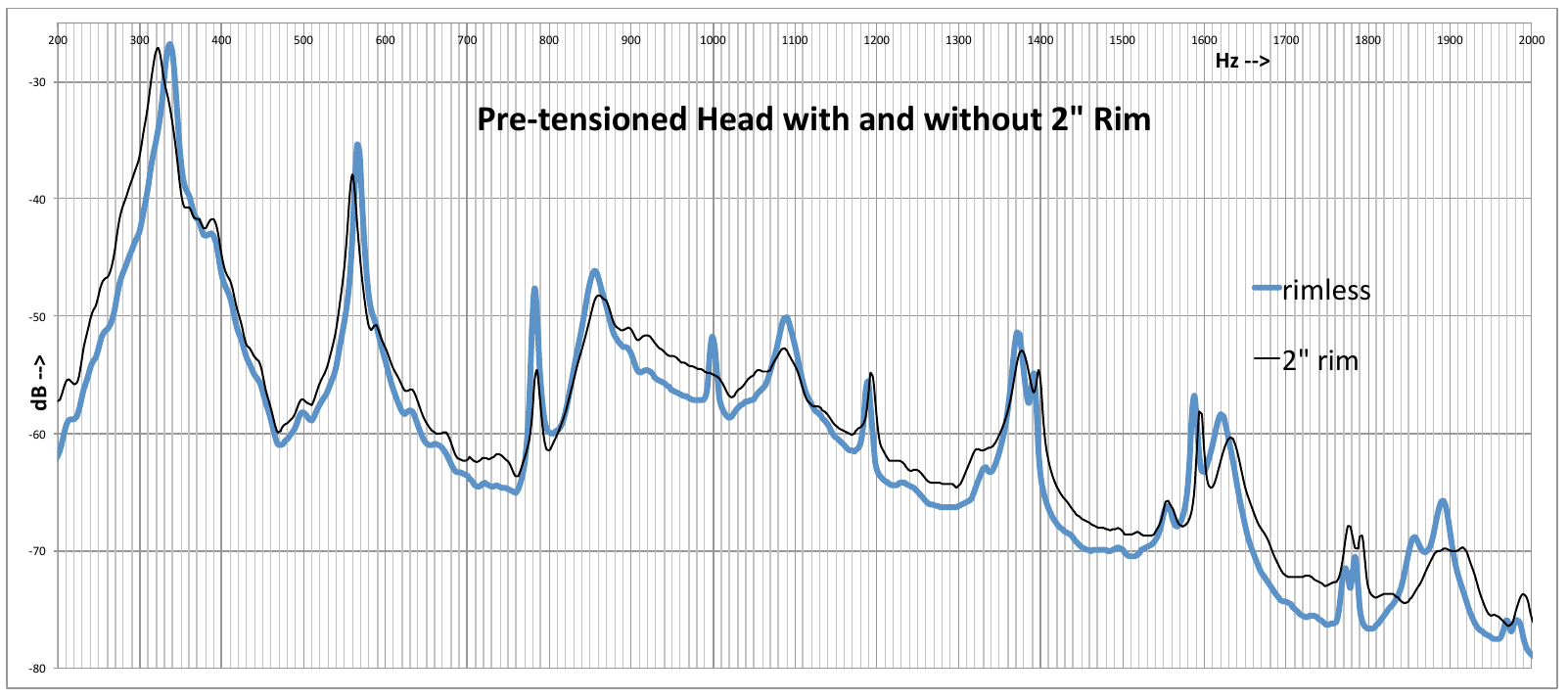}
\caption{A pre-tensioned head, with and without a $2''$ rim}
\end{figure}

Accepting the results in FIG.~4 as convincing (and with no clear alternatives), all references above and below to the $11''$ banjo head solo or alone are actually to the head mounted on a $2''$ banjo rim but with the back left completely open.

\begin{figure}[h!]
\includegraphics[width=6.0in]{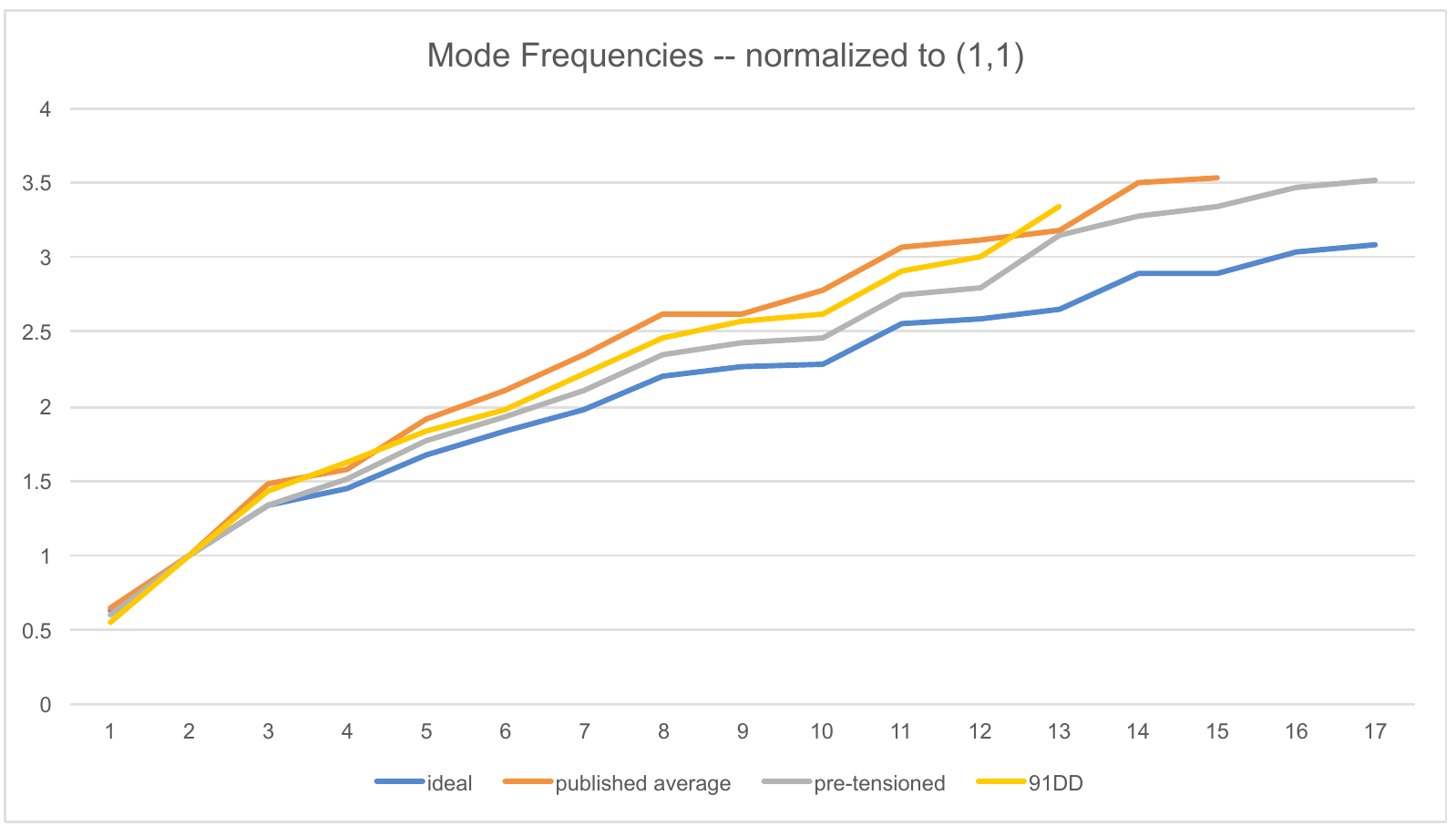}
\caption{Drum mode frequencies normalized to the (1,1) mode value {\it vs} the mode number in ascending order}
\end{figure}

FIG.~5 gives a comparison of various drum head frequency measurements.  For reference, the calculation for the ideal membrane is included.  That is the lowest frequency sequence.  In that theoretical model, all frequencies are calculated dimensionless numbers times a single reference frequency that depends of the tension and mass density of the membrane.  When considering measured values, it is reasonable to normalize the series of frequencies for a given drum head to the value of the (1,1) mode, i.e., the mode with one diameter node line and a fixed circle at its edge.  That mode is always sharper in frequency than the lowest, (0,1) mode and is easy to identify.

The sequences of measured frequencies are for 1) an average of a few published scholarly works on timpani physics, 2) the $10''$ 90DD pre-tensioned head, and 3) the $11''$ head at 91DD on a $2''$ rim with the back left wide open.  Missing from the idealized theory is the effect of the surrounding air and the stiffness of the head.  Apparently, these impact different modes to different degrees.  They do not give a single, common factor that would have disappeared when normalizing to the (1,1) mode.  And this sort of variation is to be expected.  The main value of the comparison is that the values determined in the present study by very crude means are consistent with efforts that were far more professional.

\subsection{Chadni figures}

Decades ago, laser interferometry was used to image vibrating surfaces.  Soundboards of stringed instruments were famous examples.  More recently, engineers employ a device called a ``scanning Doppler laser vibrometer."  Point it at a surface, press the right buttons, and it will tell you anything you might want to know about a vibrating surface.  Operating on a very limited budget with borrowed and rescued apparatus, I opted for Chladni figures.  Sand or some other small particle material collects at the node lines when the surface is horizontal and driven at a resonant frequency.  (Chladni, publishing in 1797, was repeating an investigation done by Robert Hooke about 100 years earlier.)

I explored the whole tension range of 85 to 91DD and tried various materials: poppy seeds, chia seeds, and sand of various sizes and origins.  I excited the head with a piezo disc double-sticky taped to the head and with a homemade, mechanical driver.  I present here only one set.  It's not the prettiest, but it was the most useful.  It is for 91DD and uses sand purchased at a crafts store.  The piezo disc produced distracting static cling; so these figures were generated using a mechanical driver.  85DD gave the sharpest and cleanest lines because it flexes and moves the most.  However, at the high end of the 100 to 2000 Hz interval, the modes get very close together.  91DD is closest to an ideal membrane because the ideal model has tension as the only important force.  The purpose of generating the Chladni figures was to match mode spatial shapes to the frequencies.  In that respect, the 91DD case followed the ideal model perfectly over the frequency range of interest.

\begin{figure}[h!]
\includegraphics[width=3.0in]{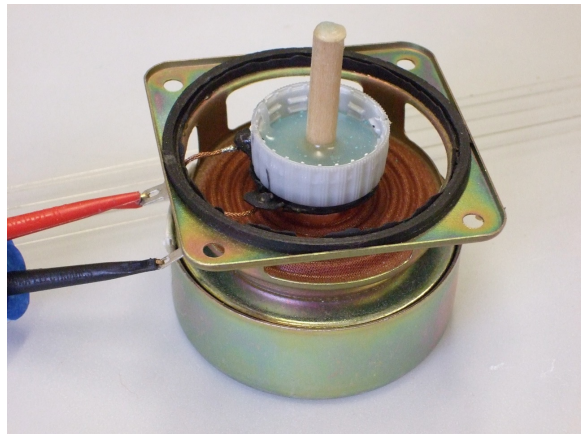}
\caption{homemade sinusoidal driver, to be attached to an amplified signal generator\cite{DIY-driver}}
\end{figure}

To generate the figures, I dialed a signal generator to each of the tap spectrum peaks, put its output through an audio amplifier, and connected that to the driver shown in FIG.~ 6.\cite{DIY-driver}  I touched the hot-glue-tipped dowel of the driver to the head.  Paying attention to the initial activity of the sand, I moved the driver to a region of greatest motion.  That clearly defined other regions reflecting the circular symmetry.  I touched the driver to the head at each of the large-motion regions and repeated until the sand figure was as sharp and symmetric as it was going to get.  Photos of those patterns are in FIG.~7.

\afterpage{\clearpage}

\begin{figure}[h!]
\includegraphics[width=5.0in]{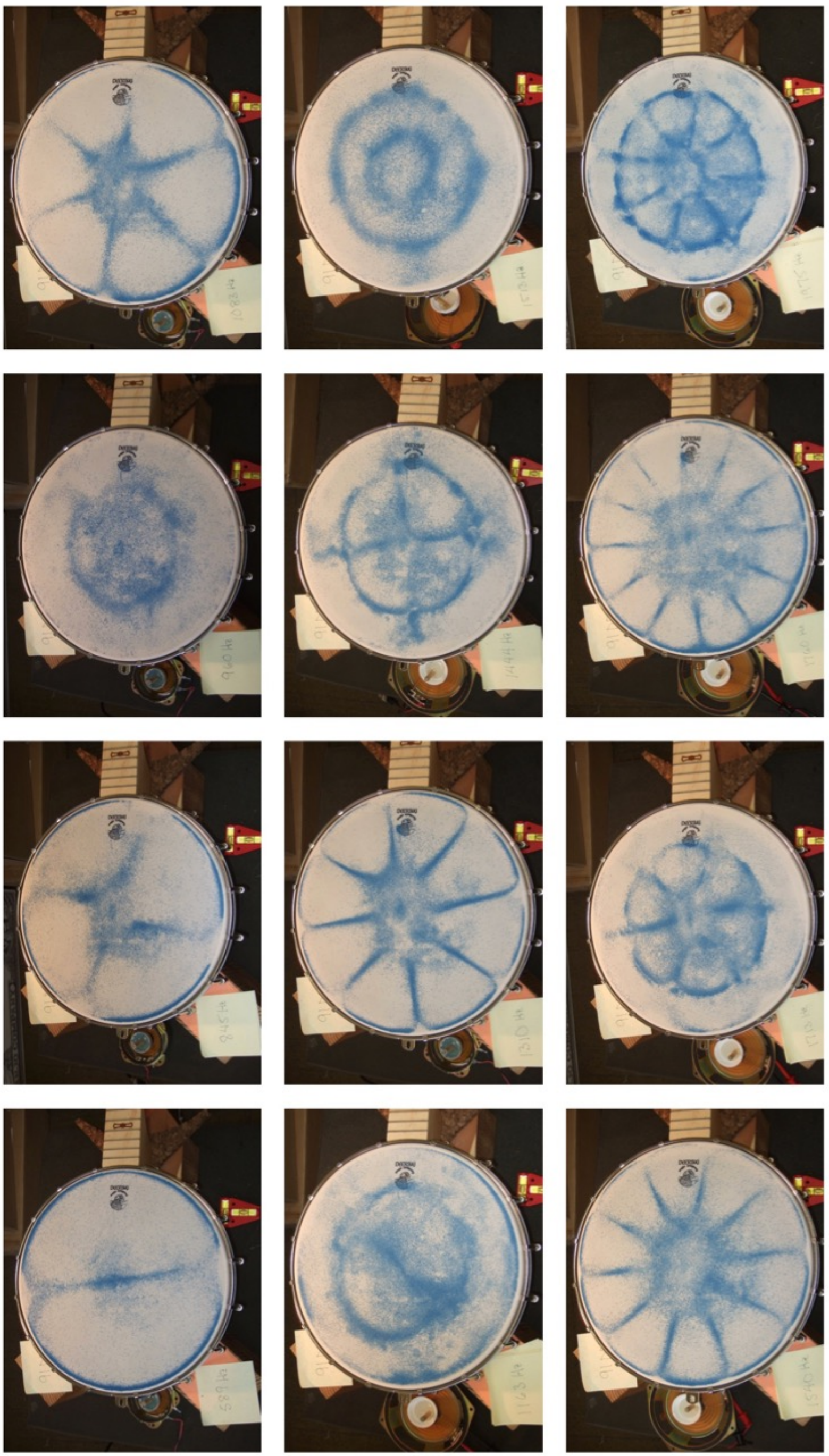}
\caption{91DD drum head Chladni figures, in order of frequency, (1,1) to (4,2), if viewed sideways. \centerline{(The (0,1) mode is not shown.)}}
\end{figure}

\afterpage{\clearpage}

FIG.~8 summarizes the relevant features of the calculations of the ideal circular membrane and the air in the ideal cylindrical cavity.  The top 12 are for the membrane.  Modes are labeled by two integers, $(n,m)$, where $n$ is the number of diameter node lines ($n=0,1,2,3...$) and $m$ is the number of circular node lines ($m=1,2,3,...$), i.e., the fixed outer edge counts as 1.  Below each circle is the mode's frequency normalized to the frequency of the lowest, $(0,1)$ mode.  (These circle node lines are not drawn carefully to scale.)

\begin{figure}[h!]
\includegraphics[width=6.5in]{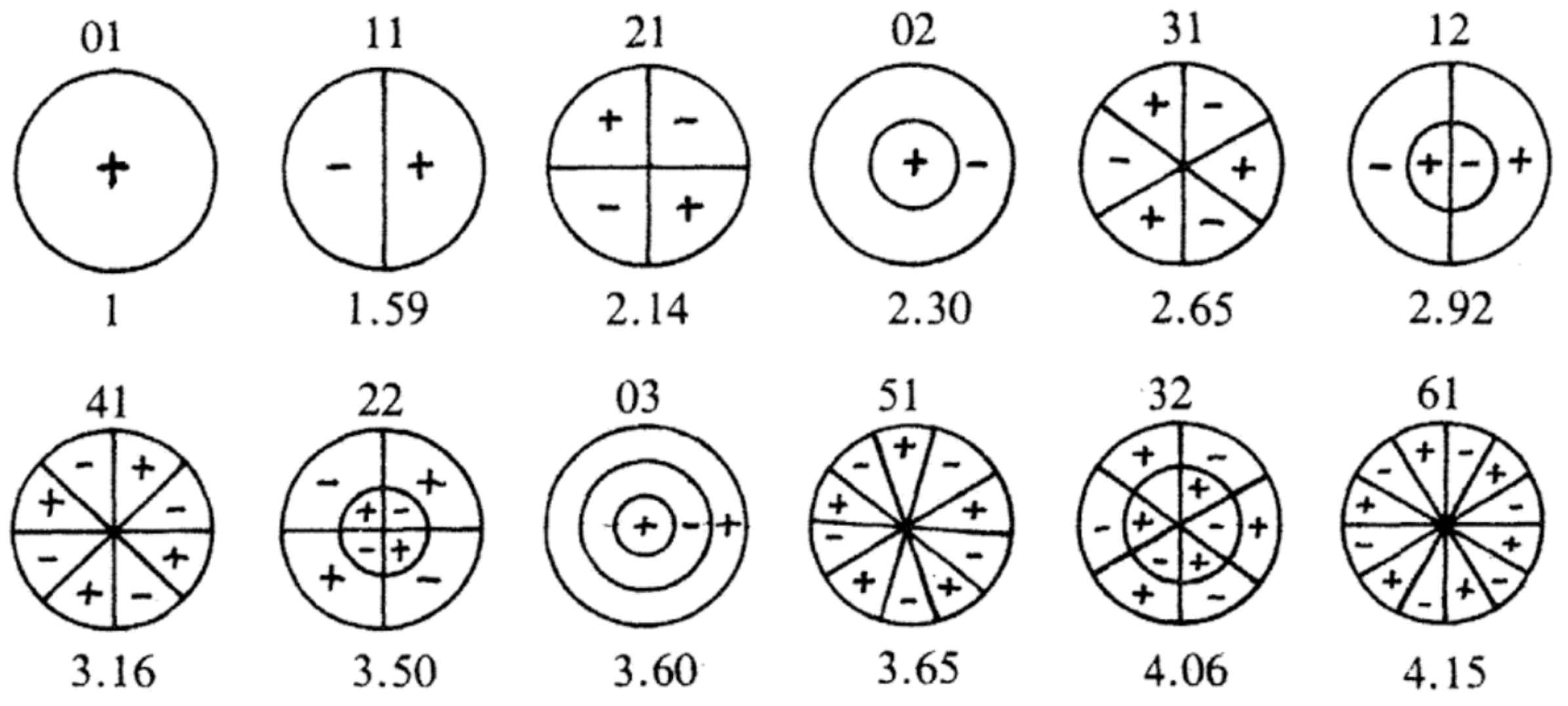}
\includegraphics[width=6.5in]{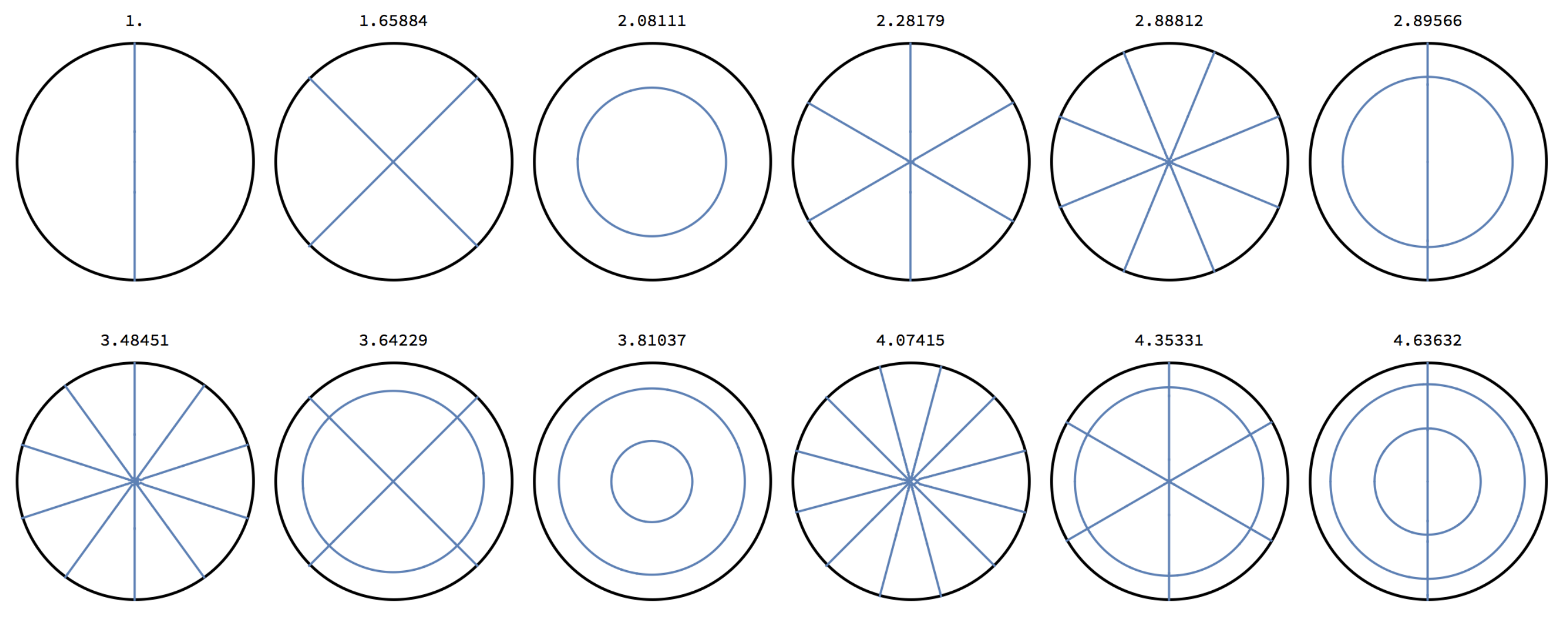}
\caption{drum head and cylinder first twelve modal node patterns --- see text regarding labels}
\end{figure}

A fine interactive graphic of the motions of the ideal membrane is on-line at 

\noindent \href{http://falstad.com/circosc/}{http://falstad.com/circosc/}.

The bottom 12 circles in FIG.~8 describe the lowest cylinder air modes in the radial and azimuthal direction.  The lines are the pressure nodes, and the outer circumference is a pressure anti-node.  In the axis direction (perpendicular to the head), there are pressure waves of integer half-wavelengths that fit top to bottom, starting with 0, i.e., constant in pressure.  (The circles here are drawn carefully with their actual radii.)  The numbers above each circle are the frequencies in units of the lowest  $(1,0)$ mode frequency.  For the dimensions of the banjos considered, only the deep pot has pressure variation in the axis direction below 2000 Hz.

\afterpage{\clearpage}

\section{Co-rod as an example of rim stiffness}

All of the work described above was done on rims with heads and no necks, strings, or bridges.  But I had previously done some measurements with the necks attached, and a tiny detail caught my eye.  The effect can be traced to having the coordinator rod (``co-rod") in place.  For these banjos, that is a ${5 \over 16}''$D metal rod firmly inside the pot that goes from the neck joint to the tailpiece.  (On older-styled banjos it is wood, known as a dowel stick.)  Clearly, it dramatically reduces rim motion along the direction of the rod's length.  

\afterpage{\clearpage}

The lowest frequency rim modes are distortions from round that go in and out in the radial direction.  The lowest such mode has four nodes around the circumference.

A very important aspect of systems with circular symmetry was not mentioned thus far because it wasn't relevant as yet and would just complicate the descriptions.  But the fact is that every mode with at least one diameter node line  is ``doubly degenerate."  That means that there are actually two distinct modes with the same frequency.  The two differ from each other by a rotation.  For the lowest rim (ring) mode with four nodes, its degenerate partner has the same shape but is rotated by $45^{\text{o}}$.  In fact, any combination (``superposition") of those two modes also has the same frequency.  That means that the motion can be positioned anywhere around the circle.  One can even arrange to have it rotate clockwise or counterclockwise.  Drum modes have the same sort of doubling.  For a given set of diameter node lines, equally spaced in angle, there is a distinct mode that is rotated relative to the first by half the angle between the original lines.  This is true for each $(n,m)$ for $n\geq 1$.  Similarly, all ring modes (i.e., with an even number $n\geq4$ of  equally spaced nodes around the ring) are doubly degenerate.

We can label positions around the rim by hours on the clock.  A co-rod placed between 12:00 and 6:00 adds stiffness to in-and-out motion between those two positions.  The originally degenerate lowest mode is now split in frequency.  The mode with nodes at 12:00--3:00--6:00--9:00 is stiffer and has a higher frequency than the one with nodes at 1:30--4:30--7:30--10:30.

\afterpage{\clearpage}

The lowest frequency head mode that will be effected by the co-rod rim stiffness is $(2,1)$, in particular the one with  node lines connecting 1:30 to 7:30 and 4:30 to 10:30.  That head mode's frequency will be raised by the additional rim stiffness because its motion pulls on the rim the most at 12:00, 3:00, 6:00, and 9:00.  The $(2,1)$ head mode rotated by $45^{\text{o}}$ will be unaffected by the co-rod stiffening.

\begin{figure}[h!]
\includegraphics[width=5.5in]{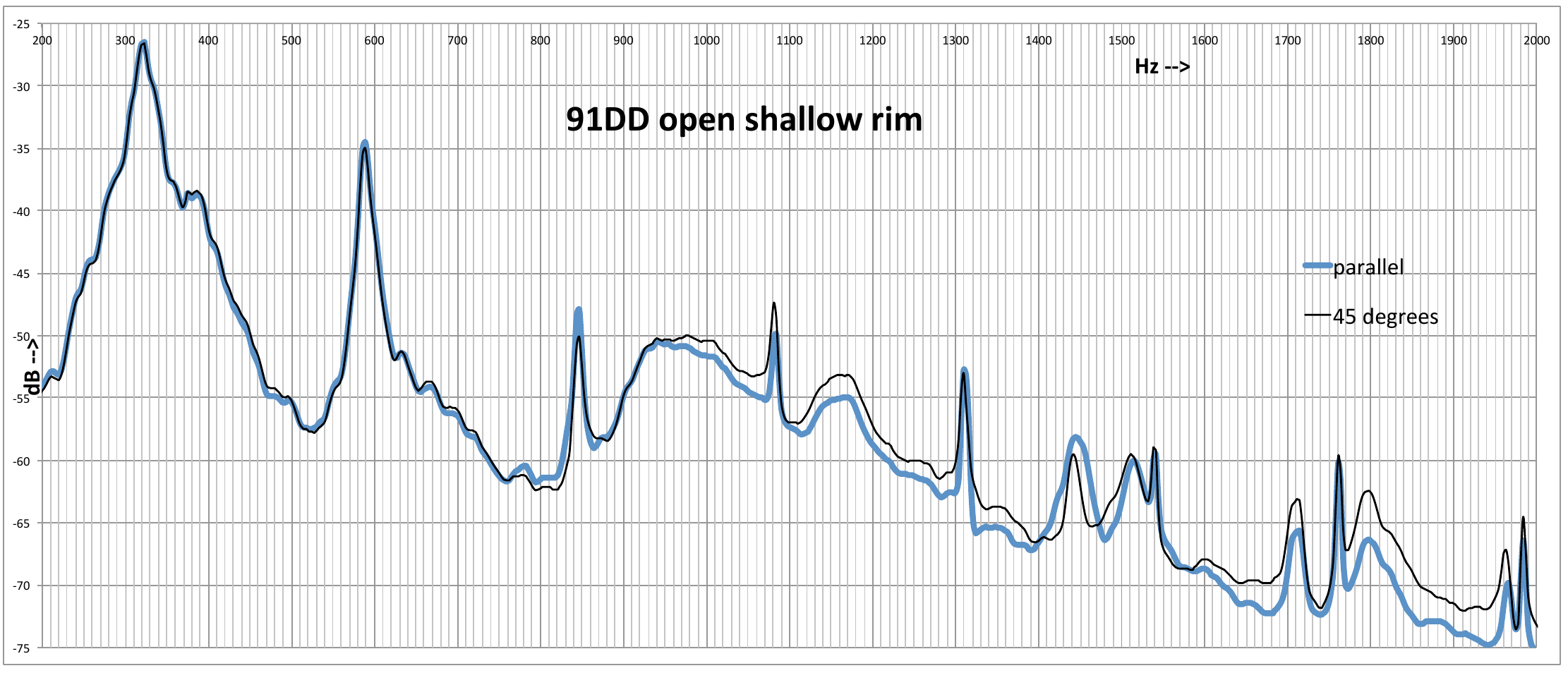}
\includegraphics[width=5.5in]{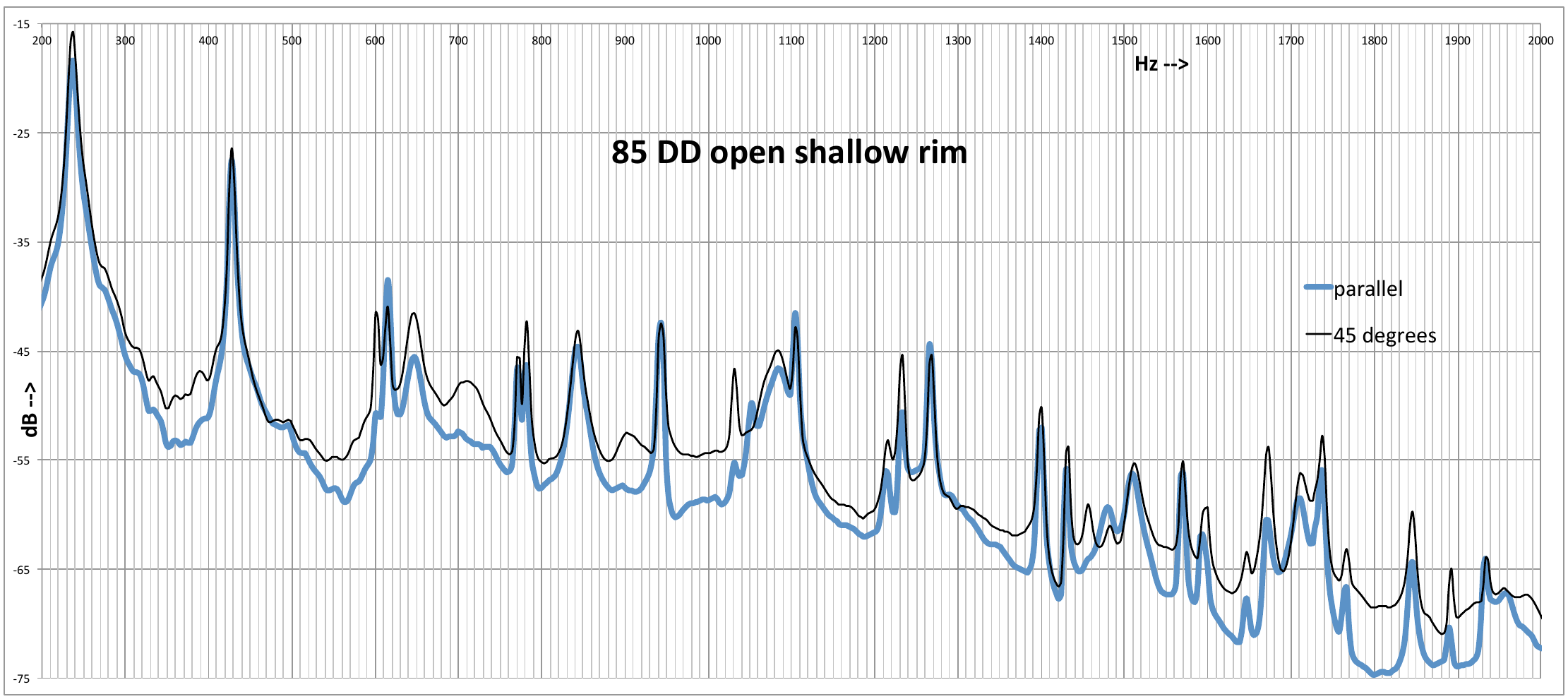}
\caption{Tap spectra for the solo head at 91 and 85 DD, comparing taps on the diameter parallel to the co-rod with taps at $45^{\text{o}}$}
\end{figure}

Head taps excite the head modes, and they produce sound.  A particular head tap is most effective at exciting a mode if it is located at one of the mode's spatial maxima.  The tap is totally ineffective at exciting a particular mode if it is located at a node.

That's the theory.  In practice, a small bit of this was actually observed.  The evidence is in FIG.~9.  At 91DD, the effect of head tension was sufficiently greater than the rim motion perturbation that no splitting was observed given the resolution of the actual experiment.  However, at 85DD, the $(2,1)$ mode frequency splitting is quite evident.  It behaves qualitatively as expected.  Taps along a diameter line parallel to the co-rod are more effective at exciting the higher of the split pair.  Taps along a line rotated by $45^{\text{o}}$ show a relatively much stronger excitation of the lower mode.  Indeed, tapping $\sim$400 times in rapid sequence with a hand-held piano hammer does not provide precise locating.  The taps were made all along the diameter line; so their angular positioning was less accurate as they neared the center.  Nevertheless, the effect, with the right qualitative features, is certainly there.

Other modes are also seen to be split for 85DD and not 91DD.  However, the chosen two lines of taps, i.e., at $45^{\text{o}}$ with respect to each other, are not ideal or are even totally ineffective at preferentially exciting one of a pair {\it versus} the other.

\section{more practical details}

\subsection{Mic position, room sound, \& floor bounce}

FIG.s~10--12 illustrate some perils of sound recording.  Actually, these same issues impact any sound in enclosed spaces, whether recorded by microphone or heard by ear.  However, they are exaggerated here by restricting the sound production to a single frequency at a time and by listening with a single small microphone.  The issues arise when the detected sound is a sum of a direct component as produced by the source and reflected components.  All have the same frequency.  Their relative phase as they arrive at the detector dramatically effects how they combine to be either stronger or weaker than the direct sound itself.

These blue-on-grey figures are screen shots of recordings made with Audacity.  The vertical axis is microphone voltage.  And the horizontal axis is time.

FIG.s~10, 11, and 12 all are $3{1 \over 2}$ minute recordings of sound from a $2{1 \over 4}''$ speaker driven by a signal generator and audio amp set to scan linearly from 100 to 2000 Hz.  The scan is slow enough that the changes in frequency over the time-of-arrival differences from various reflections in the room are negligible.

\begin{figure}[h!]
\includegraphics[width=5.5in]{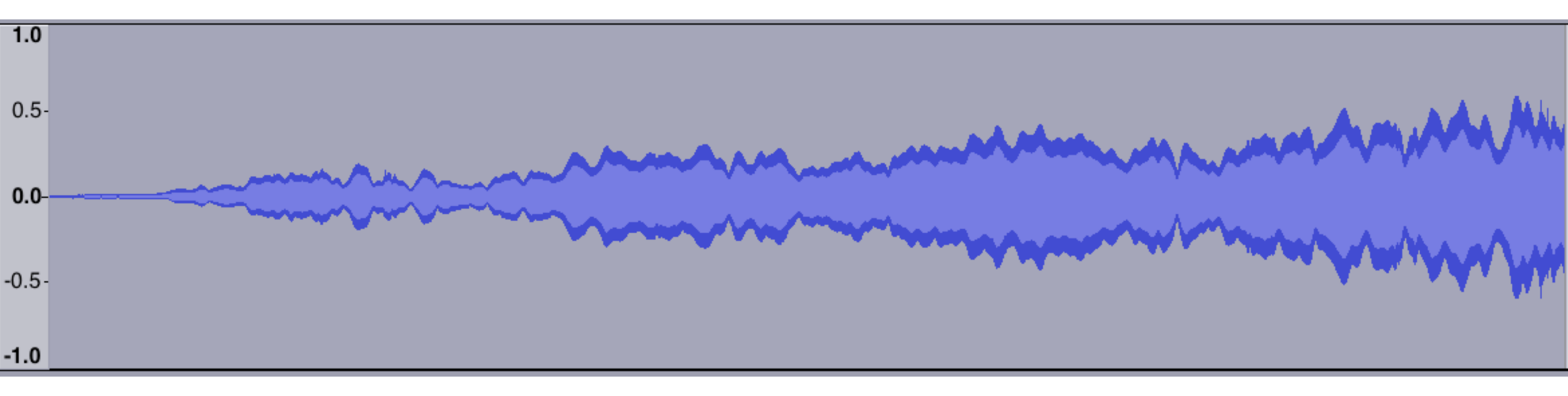}
\includegraphics[width=5.5in]{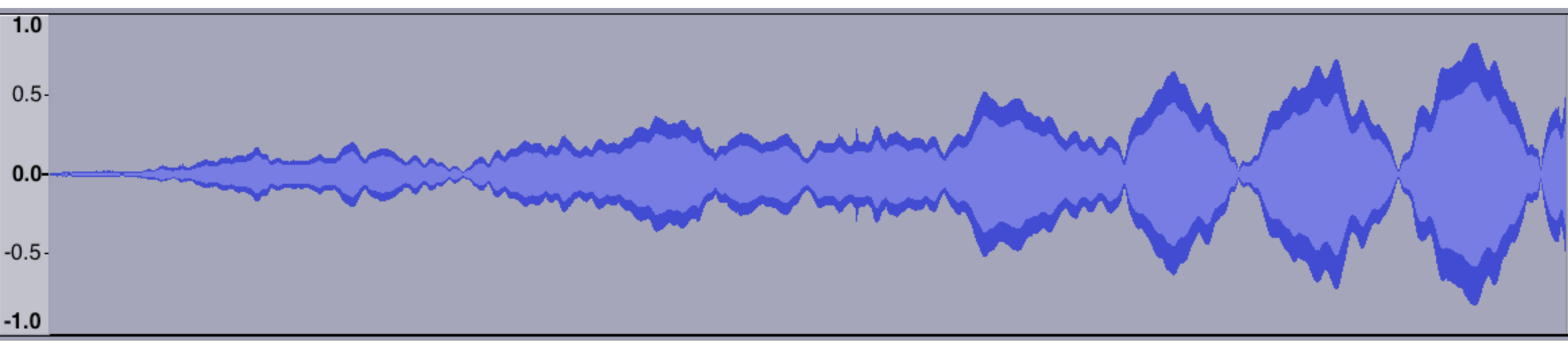}
\caption{microphone voltage {\it vs} time ($3{1 \over 2}$ min total) 100 to 2000 Hz linear scans with constant driving voltage amplitude, recorded at 3ft, at two different places in the room}
\end{figure}

In FIG.~10 the microphone faces the speaker and is placed 3 feet away.  The two plots correspond to placing the speaker -- microphone pair at two different places in the room, oriented in different directions relative to the walls.

FIG.~11 shows the recording of the same speaker and drive and the same microphone at the same distance, but this time through a $4''$D cardboard tube --- as shown in FIG.~13.

\begin{figure}[h!]
\includegraphics[width=5.5in]{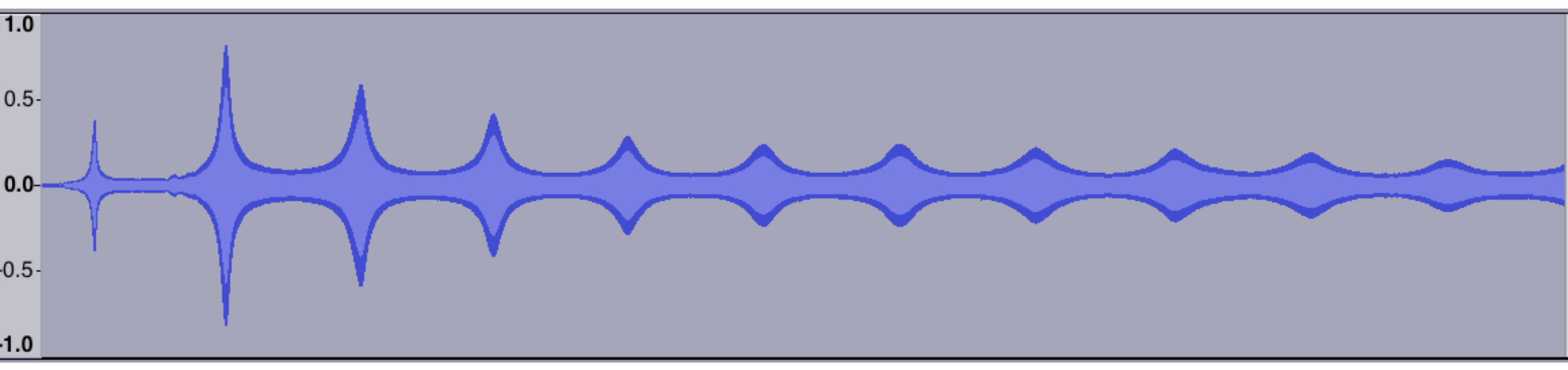}
\caption{microphone voltage {\it vs} time ($3{1 \over 2}$ min total) 100 to 2000 Hz linear scan with constant driving voltage amplitude, recorded through a 3 ft, 4 in diameter cardboard tube}
\end{figure}

FIG.~12 shows the recording of the same speaker and drive.  This time the microphone is only 1 foot away.  More importantly, both the speaker and microphone are on the carpeted floor.

\begin{figure}[h!]
\includegraphics[width=5.5in]{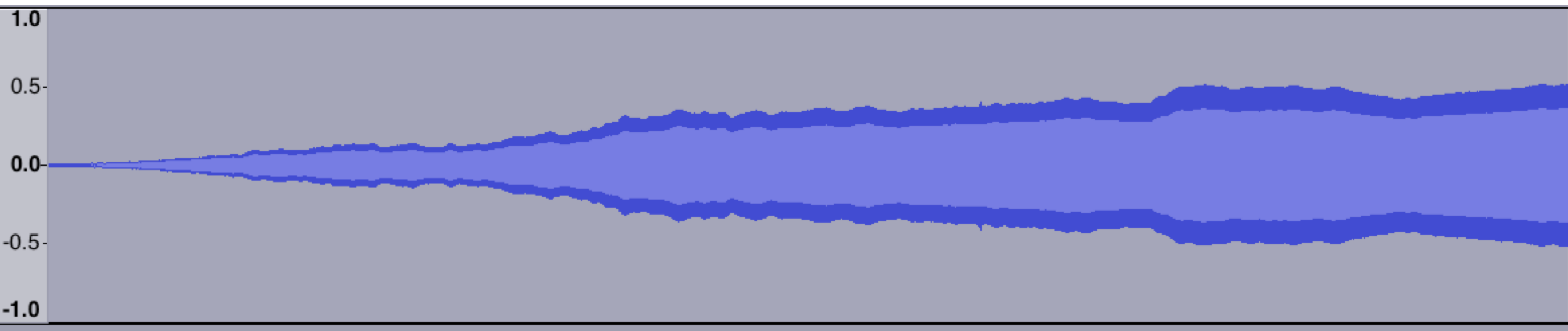}
\caption{microphone voltage {\it vs} time ($3{1 \over 2}$ min total) 100 to 2000 Hz linear scan with constant driving voltage amplitude, recorded at 1 ft, with both speaker and microphone on the floor}
\end{figure}

\begin{figure}[h!]
\includegraphics[width=3.7in]{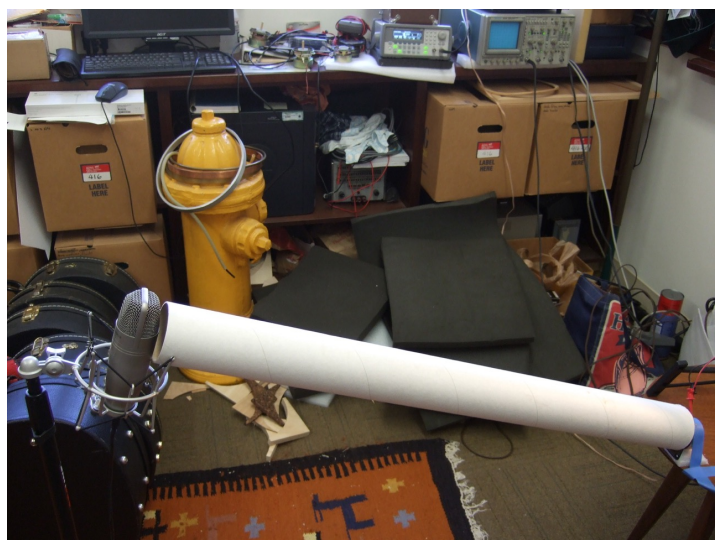}
\caption{recording through a 3 ft, 4 in diameter cardboard tube}
\end{figure}

The motivation for all these shenanigans was to understand which peaks in my careful instrument recordings were directly attributable to the instrument and which had to do with the interaction with the room.

FIG.~11 is a dramatic example of the basic physics of wind instruments.  The source generates (or at least has the potential to generate) sounds at a great variety of frequencies.  The sound travels down the tube and reflects off the other end (and then back again, too).  The large amplitudes arise when the frequency and tube length put the reflected wave in phase with the direct wave (and the other reflected waves) at the microphone.  (You could use FIG.~11 and the data given [the length might be closer to $37''$] to estimate the speed of sound.)

The two examples in FIG.~10 show how much reflections off objects and walls can impact the total sound at the microphone.  Altering the position of the speaker and mic with respect to the room totally alters how those reflected waves combine.

Moving the mic closer to the speaker, as for FIG.~12 decreases the amplitude ratio of the reflected sounds relative to the sound that arrives at the mic directly from the speaker.  That is because each reflection also involves some dissipation of the wave energy.  Putting the speaker and mic on the floor has an additional benefit.  All conceivable paths for reflected waves are much longer than the direct path.  This reduces ``floor bounce," a phenomenon known to recording engineers, due to the direct sound being effected noticeably by sound that bounces off the floor roughly half way between the source and the mic.  The small speaker is not efficient at low frequencies.  So, even though the electrical driving amplitude is constant throughout the frequency sweep, the generated sound increases in amplitude as the frequency increases.

At these frequencies the room resonances are not separately distinguishable.  Dissipation gives an isolated resonance a width in frequency.  At these frequencies and without dissipation, the room would have many resonances {\it very} close together.  Their actual widths smear the whole thing out.  (It's not a stall shower.)  Rather, a specific frequency steady direct sound sets up a room sound field whose amplitude is static.  There will be louder and softer places and possibly some actual nodal planes.  If the frequency changes very slowly, this pattern will shift slowly in space.  In particular, nodal planes will pass through the fixed microphone as the frequency slowly varies.  That is what's responsible for the dramatic amplitude variations in FIG.~10.  Another way to demonstrate this is the recording plotted in FIG.~13.  It uses the same speaker and microphone.  This time the frequency is {\bf fixed} at 1200 Hz.  It runs for 55 seconds while the microphone is moved as smoothly as I possibly could by hand.  The motion is from outstretched arm on one side to outstretched arm on the other while standing 5 feet from the speaker.  So the microphone is moving through the rather complicated field of maxima and minima of the static 1200 Hz sound set up by the fixed speaker.  (The tiny, short bits of noise come from handling the microphone in the process.)

\begin{figure}[h!]
\includegraphics[width=5.5in]{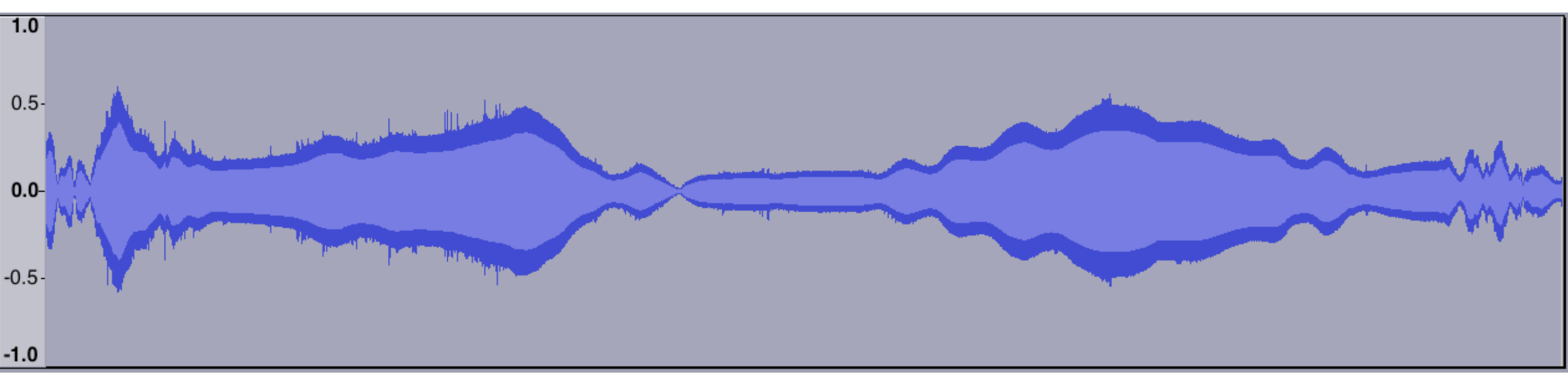}
\caption{microphone voltage {\it vs} time (55 sec total) at fixed amplitude 1200 Hz , standing 5 ft away, moving the microphone horizontally about 6 ft, i.e., from outstretched arm on one side to the other}
\end{figure}

\subsection{Head taps}

 \subsubsection{relevance to bridge position}

\href{http://www.its.caltech.edu/~politzer/air-head-exp/open-shallow-9-across.mp3}{http://www.its.caltech.edu/$\sim$politzer/air-head-exp/open-shallow-9-across.mp3} is a recording of nine head taps, spaced more-or-less evenly across the head diameter.  A tap {\it exactly} at the center can only excite the $(0,n)$ head modes and misses out on most of the banjo's potential.  Refer to FIG.~8.  An actual bridge would straddle the center and excite, albeit weakly, many azimuthal modes as it rocked.

\subsubsection{head taps: reproducibility}

The are several ways to produce sound which can be analyzed for its frequency spectrum.  For the analysis presented here, I opted for a long sequence of head taps with a piano hammer.  Except where otherwise noted, it's about 400 taps distributed uniformly around on the head.  A reasonable alternative is a piezo disk, stuck to the head, driven by a signal generator.  In that case, the strength of various peaks would depend strongly on the position of the piezo --- and some would be absent if they had a node there.  For a different study, a piezo tucked between the foot of the bridge (with tensioned strings) and the head would certainly be of interest.  That, after all, is where the strings excite the head, and the resulting sound is certainly depends on that position.

\begin{figure}[h!]
\includegraphics[width=3.7in]{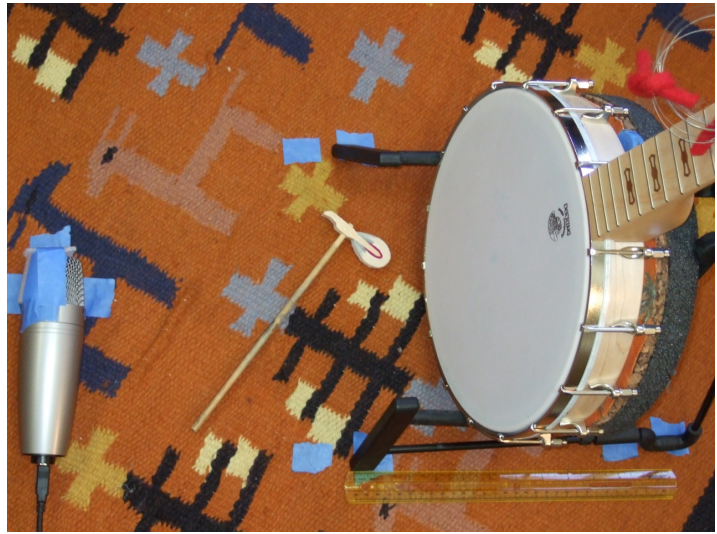}
\caption{microphone on the floor to record head taps on the shallow rim with the belly back}
\end{figure}

I also took the room sound concern to heart and ended up recording with mic and banjo on the floor, separated by  about $18''$.  For the various rims and backs, I made an effort to keep the head--mic spacing and position relative to the room the same from one to another.

FIG.~16 illustrates the extent to which the resulting spectra are reproducible.  The lower two lines are two consecutive $\sim$400 tap runs on the same set-up, separated by a few minutes.  Whatever determined the positions of the peaks did the same in both runs.  The upper two curves are displaced together by a fixed amount in dB from the lower two to make the relevant comparisons more visible.  Those two runs were separated by a couple of weeks.  Masking tape marked the approximate floor position over that period, but objects were moved in the room.  The biggest difference between the two curves is a systematic fractional shift of the peak frequencies, implying a slight difference in tension.  I regard the magnitude as well within the variation in DrumDial reading and the possible settling of a head in the hours after last checking.  (On many days, the head tension was readjusted a few times rather than letting it sit and stabilize for days.)

\begin{figure}[h!]
\includegraphics[width=5.5in]{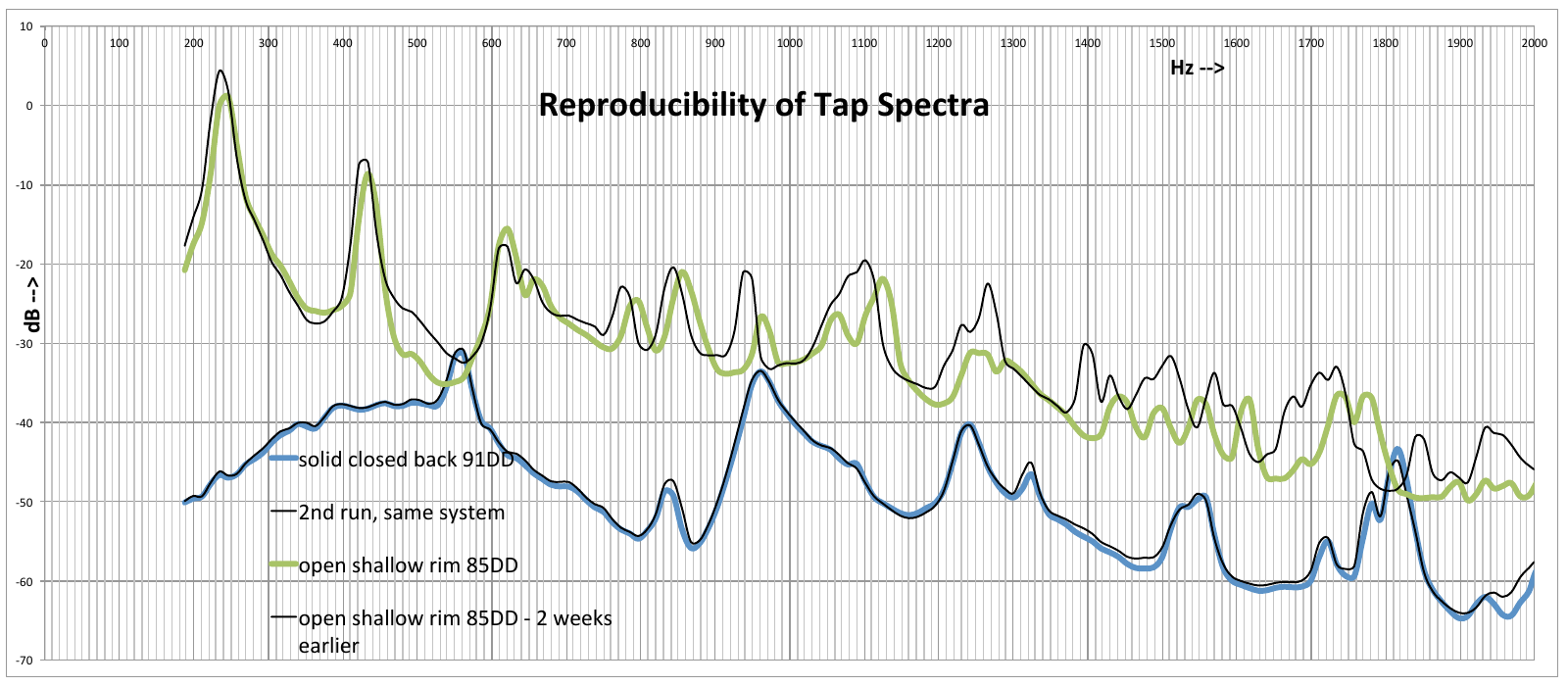}
\caption{Reproducibility of tap spectra}
\end{figure}

\subsection{The DrumDial}

The commercial Drumdial has a spring-loaded plunger that can depress the drum head relative to the outer circumference of the DrumDial bottom when it is placed on the head.  The magnitude of that depression shows up on the dial.

\begin{figure}[h!]
\includegraphics[width=3.0in]{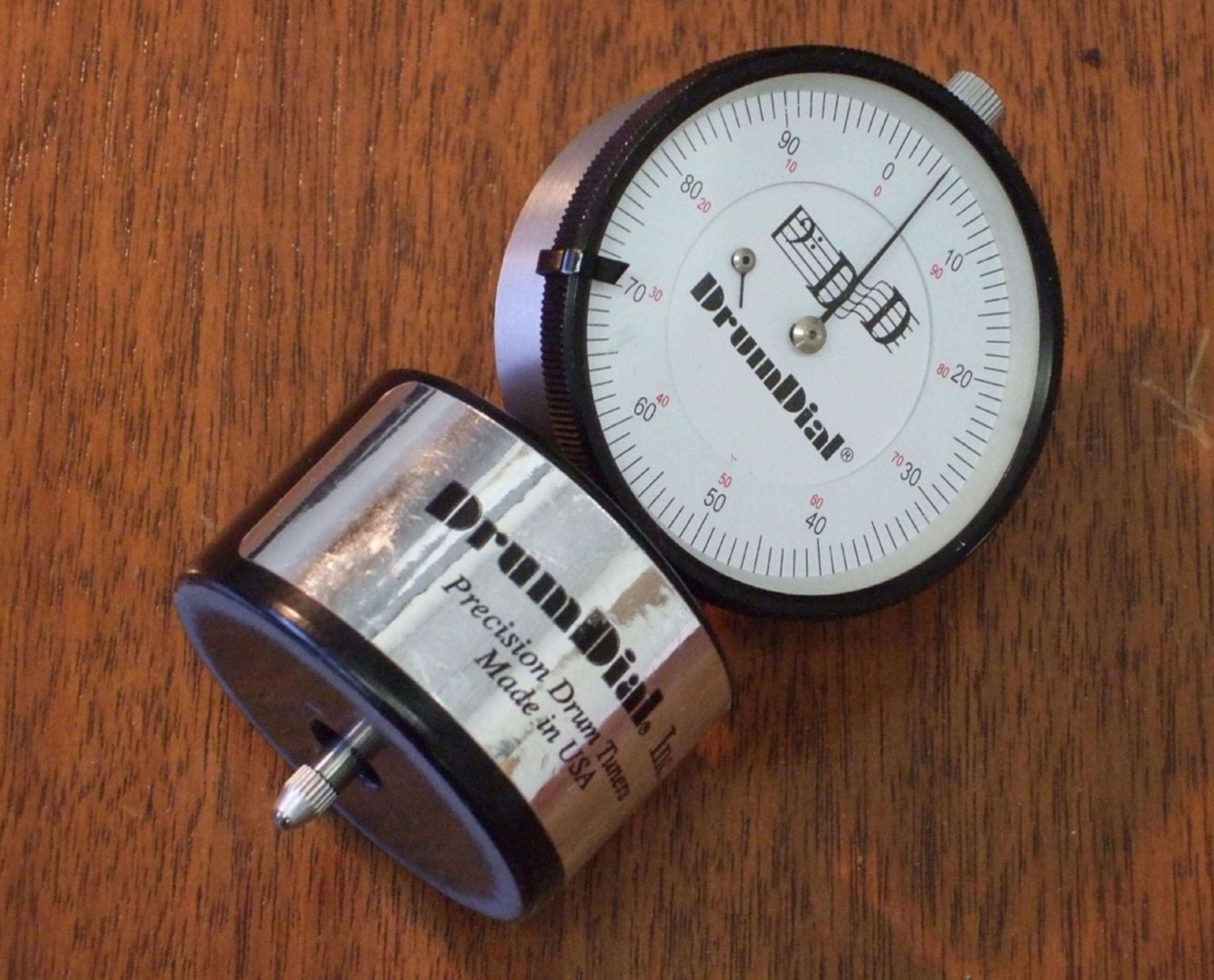}
\caption{a standard DrumDial}
\end{figure}

I wrote to the DrumDial Corp., and they cheerfully sent me their version of a conversion of dial reading to tension.  Delighted to get a response, I did not ask further for methodology, assumptions, or uncertainties.  The numbers agree roughly with a previous banjo estimate.\cite{dickey}  The dots in FIG.~18 are the numbers I received.  The solid line is my own, crude, smooth fit that takes into account the fact that 100 on the dial corresponds to infinite tension, and the tension $\cal T$ for dial reading ${\cal D}$ should go like ${\cal T} \propto 1/(100 - {\cal D})$.  This is actually of some use in the present investigations because the predicted frequencies for an ideal membrane scale with square root of the tension.  For example, the calibration curve suggests that the peak frequency shift for 85 down from 91 DD is about a factor of 0.74.
\begin{figure}[h!]
\includegraphics[width=5.5in]{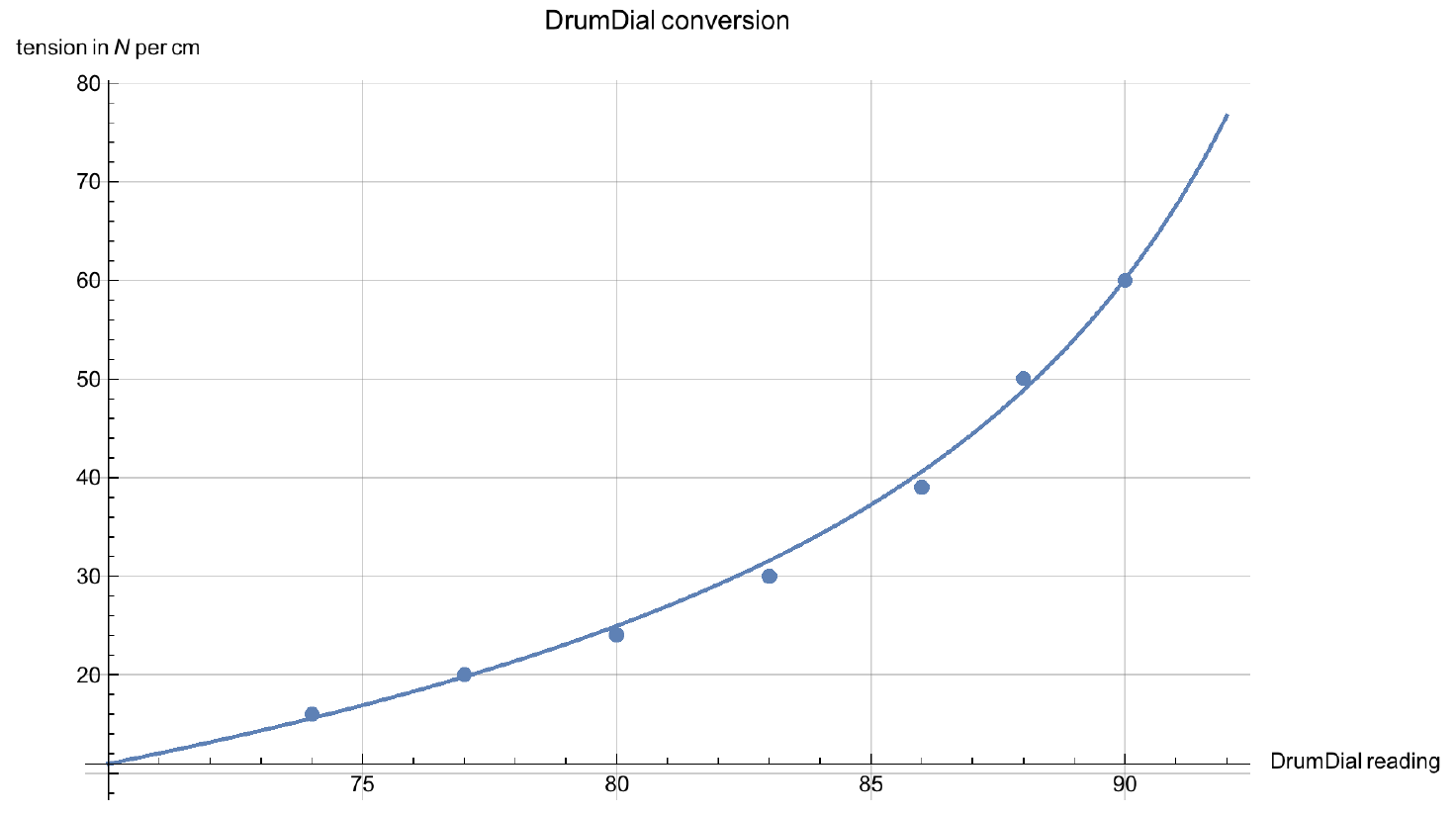}
\caption{DrumDial--tension conversion; DrumDial, Inc. data \& a reasonable, smooth fit}
\end{figure}

\newpage

\subsection{a test of Audacity}

I used a randomly chosen sound file (actually the one that appears in the second plot of FIG.~10) to demonstrate that Audacity's spectrum calculations are likely correct.  FIG.~19 has two Audacity screen shots that relate to the same sound recording.  The upper one is microphone voltage versus time (0 to $3{1 \over2}$ minutes).  The vertical voltage scale is chosen to be in decibels, (i.e., logarithmic and not linear).  The lower scale is Audacity's calculation of the frequency spectrum for that entire recording.  The horizontal scale is frequency in Hertz.  The vertical scale is the spectral amplitude on a decibel scale.

\begin{figure}[h!]
\includegraphics[width=4.5in]{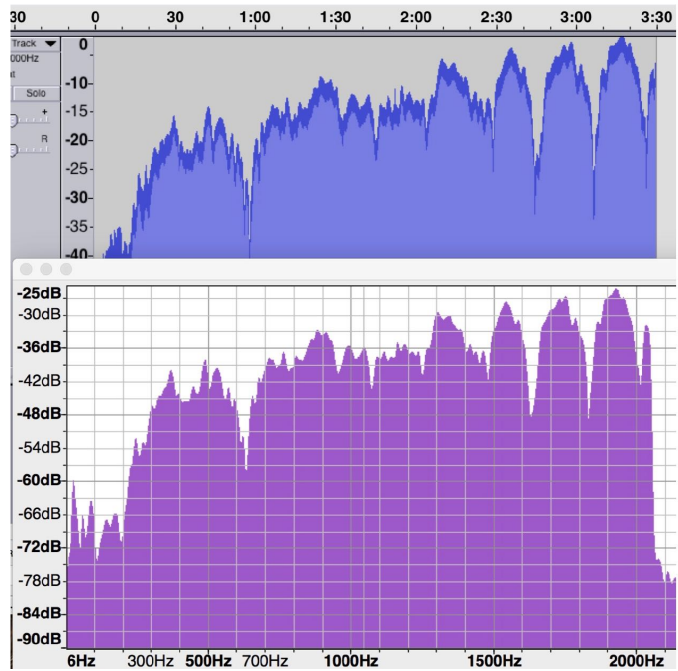}
\caption{A single sound file plotted {\it vs} time and spectrum analyzed analyzed by Audacity }
\end{figure}

The speaker drive was a scan, linear in frequency, from 100 to 2000 Hz performed over the $3{1 \over2}$ minutes.  I rescaled the images (i.e., not changing any Audacity setting) so that 100 Hz on the spectrum fell directly under the beginning of the scan in time and 2000 Hz fell directly under the end.  Audacity allowed some adjustment of the vertical displays, and I did that until the variations in the two graphs looked similar.  Scanning frequencies linearly in time gives a linear relation between frequency and time for the recording.

\section{Another way?}

{\small \it Focus once more: What is the goal here?  For me, it is to connect some simple aspect of physics to some recognizable feature of sound.  My earliest motivation was learning the astonishing account of the difference between oboe and clarinet.  Double-{\it versus}-single reed turns out to be an effect, not a cause.  The main contrast is conical-{\it versus}-straight bore.  And it's impossible to appreciate how astounding that is from the simple derivation of the harmonic spectrum for each case.  You have to listen to a computer programed to play a sound with a very rich spectrum of integer multiples of some fundamental frequency.  And compare that to the sound produced by leaving out all the even multiples.  No reeds, no ombouchure, no pitch holes, no exotic wood... ...Inspired by John Chowning, I tried to do something like that for the banjo.  But it's all still quite debatable.}

Is the problem the limitation to frequencies under 2000 Hz?  Here is a crude look at tap spectra going up to 10,000 Hz  (i.e., FIG.~20.)  Of course, there are differences.  And no explanations come to mind.  Actually, I am surprised by the similarity.

\begin{figure}[h!]
\includegraphics[width=5.5in]{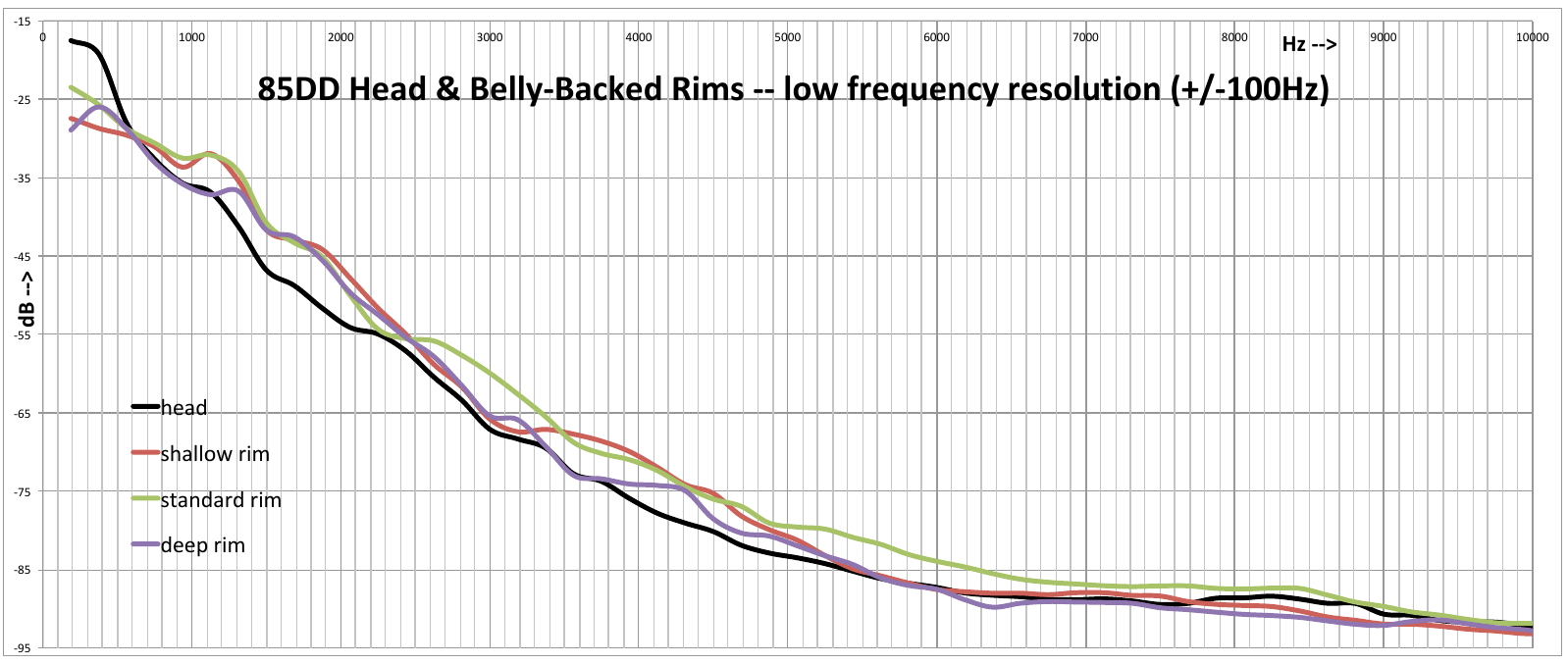}
\caption{head and three backed rim tap spectra at low resolution ($\pm 100 Hz$) at 85DD}
\end{figure}

Frequency analysis, normal modes, resonant frequencies, \&c. are very useful, well-developed concepts in many areas of physics.  The ``measured" spectra presented here are evaluated for some entire time interval, e.g., the duration of a tap sound (and then repeated multiple times to get a stable, higher resolution spectrum).  But the spectrum does not contain all the information of the recorded sound.   Yes, there is an information-preserving transformation from sound-in-time to sound-in-frequency.  However, the frequency domain description packs very important features into phase relations between different frequencies.  A single clap of thunder and the sound of a waterfall can have very similar frequency spectra.  They certainly differ in their time dependence.  In terms of a frequency description, that difference goes into the phases, which are not mentioned when talking spectra.

A spectrogram is an attempt to get timing information into a frequency analysis.  For example, FIG.~21 is a spectrogram of \href{http://www.its.caltech.edu/~politzer/air-head-exp/open-shallow-9-across.mp3}{the nine taps taken in order across a diameter of the head, i.e., of the recording featured in \S VII.B.1.}   It attempts to convey the spectral content as a function of time. 

\begin{figure}[h!]
\includegraphics[width=5.5in]{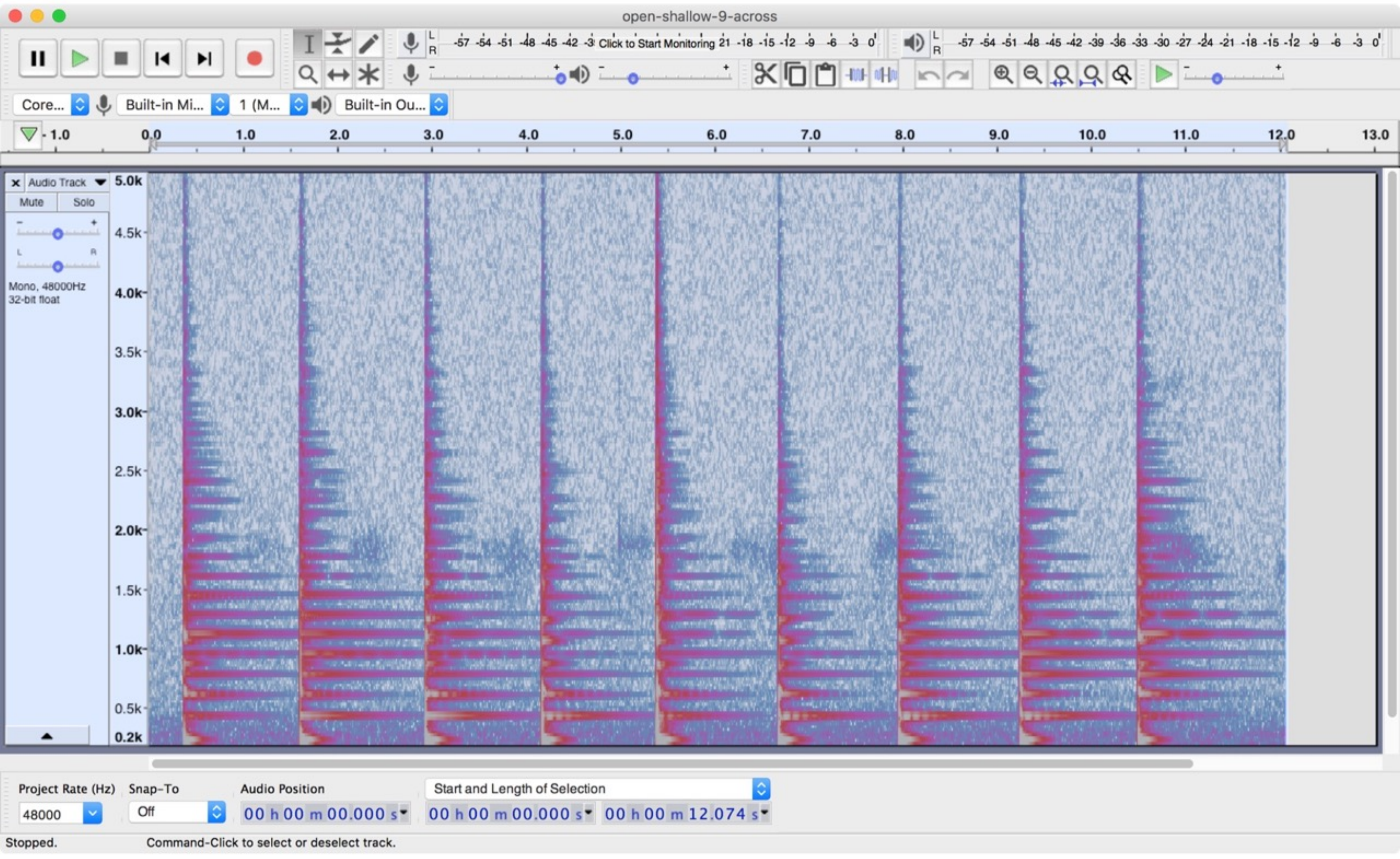}
\caption{Spectrogram of the recorded 9 taps across a diameter, as appears in \S VII.B.1}
\end{figure}

But it involves a trade-off that loses information, rather like the spectrum analysis but just a different choice of what's kept and what's sacrificed.  Audacity's Spectrogram Settings panel reflects this inevitability.  In addition to choosing whether to plot frequencies on a linear or a logarithmic scale, you are offered seven other choices, each with a range of possible numerical values.  And the choices produce spectrograms that highlight different features.  They can look very different for the same sound recording.  It's just a reflection of the wave or Fourier transform  ``uncertainty principle."  There is a necessary compromise between frequency precision and timing precision.

For FIG.~21, I chose spectrogram settings to emphasize what I found interesting.   The tap near the center (both the middle of the nine and the middle of the head) sounds like a thud.  It doesn't ring at all.  Nevertheless, it has more power or strength at the highest frequencies shown than any of the other taps.  In fact, that is what combines to create the thud.  The ringing quality of taps nearer the rim is likely reflected in the 2000 to 3500 Hz power and its persistence in time.

What's needed is a better understanding of the relation of the impressions of musical sound to measurable quantities.  This endeavor is called ``psychoacoustics," and it's an old, established, and rich field of inquiry.  But I don't think they've found the answers as yet.

\bigskip

{\it \footnotesize \{The order of the four recorded versions is 85shallow, 85deep, 91deep, 91shallow.\}}


\begin{thebibliography}{99}
\bibitem{DIY-driver}For the DIY acoustician, a mechanical oscillator as in FIG.~6 can be driven with a lap-top computer using Audacity.  Under Generate $\rightarrow$ Chirp, you can design a frequency sweep, and under Generate $\rightarrow$ Tone you can dial in a particular frequency.  The external speaker output is sufficient to drive that oscillator.
\end{thebibliography}
\end{document}